# Genetically encoding stimulated Raman-scattering probes for cell imaging using infrared fluorescent proteins

David Regan[1]‡, Ozan Aksakal[1]‡, Athena Zitti[1], John McLarnon[1], Magdalena Lipka-Lloyd[2], Pierre J. Rizkallah[2], Anna J. Warren[3], Peter D. Watson[1], Wolfgang Langbein[4*], D. Dafydd Jones[1*], Paola Borri[1*]

1. Molecular Bioscience Division, School of Biosciences, Cardiff University, Cardiff, UK. 2. School of Medicine, Cardiff University, Cardiff, UK. 3. Diamond Light Source Ltd, Harwell Science and Innovation Campus, Harwell, UK. 4. School of Physics and Astronomy, Cardiff University, Cardiff, UK.

**ABSTRACT:** Stimulated Raman scattering (SRS) microscopy offers great potential to surpass fluorescent-based approaches, owing to the sharp linewidth of Raman vibrations amenable to super-multiplex cell imaging, but currently lacks one crucial component: genetically encodable tags equivalent to fluorescent proteins. Here, we show that infrared fluorescent proteins (IRFPs) can be used as genetically encoded SRS probes and benefit from the electronic pre-resonant SRS enhancement effect with near-infrared exciting pulses, comparable to synthetic dyes reported in the literature. SRS imaging of the nucleus in mammalian cells is demonstrated where a histone protein is fused to an IRFP. This work opens the route towards Raman-based cell imaging using genetically encoded probes, motivating efforts in solving the challenges of photostability and creating a vibrational palette.

The ability to study complex biomolecular events *in situ* using optical microscopy approaches has revolutionised our understanding of fundamental processes essential to life and disease. The underlying paradigm shift was coupling fluorescence microscopy with probes such as the green fluorescent protein (GFP)[1–3] that allowed generation of genetic fusions, whereby only specific molecular targets are labelled directly within the cell. Beyond GFP, many fluorescent proteins (FPs) spanning the near ultraviolet, visible and near infra-red regions have been developed, expanding the utility of fluorescence microscopy in cell imaging. These include infrared fluorescent proteins, such as mRhubarb720[4] emIRFP670[5] and mIRFP670nano3[6] that use the haem breakdown product biliverdin Xi-α (BV) as the fluorescent cofactor (Fig. 1a). Far-red FPs such as mCherry[7,8] have the chromophore directly encoded within the amino acid sequence and formed through a post-translational covalent rearrangement event (Fig. 1b).

Fluorescence microscopy using FPs is the current method of choice for cell imaging, supported by its high specificity and sensitivity. However, some limitations remain. Organic fluorophores, such as those present in FPs, are prone to photobleaching, limiting time course observations and quantitative analysis, and can be cytotoxic.[9–12] IRFPs also have generally poor quantum yields (<15%).[5,13] Moreover, fluorescence excitation and emission spectra are intrinsically broad (typically 50 to 100 nm in wavelength or 1000 to 2000 $cm^{-1}$ in wavenumber) and featureless. This generates a "colour barrier" that restricts the number of distinguishable fluorescent probes, and corresponding biomolecular targets, typically to about 5 probes per cell. The latter limit is especially important, as the ever-expanding complexity of biological systems requires the simultaneous monitoring of a multitude of biochemical components (often more than 20), and these are impossible to track simultaneously with current fluorescence techniques.

Vibrational microscopy based on Raman scattering provides an alternative approach with the potential to overcome the above-mentioned limitations of fluorescence.[14] The inelastic Raman scattering of light by vibrating chemical bonds is photostable, and vibrational resonances of most biomolecules in liquids are spectrally narrow (below 20 $cm^{-1}$). However, a major limitation of spontaneous Raman scattering is that photon fluxes in detection are extremely low, due to the small Raman scattering cross-sections of typical chemical vibrations. As a result, Raman microscopy requires long integration times and/or large incident powers, often incompatible with imaging living cells and tissues. Furthermore, fluorescence backgrounds can easily overwhelm the Raman signal. To overcome these limitations several approaches have been proposed.[15] One of the most exciting methods, electronically pre-resonant coherent Raman scattering (epr-CRS)[16] combines the sensitivity of electronically-resonant Raman scattering with the coherent enhancement of nonlinear CRS.[17,18] In CRS, two pulsed laser beams, called pump and Stokes, are used to coherently drive molecular vibrations. As a result, identical chemical bonds in the focal volume are driven in sync and Raman scattered light constructively interferes, generating coherent signal enhancement. Typical CRS setups use near-IR laser sources to achieve deep penetration while minimising multiphoton damage. Hence epr-CRS requires a far-red/near-IR chromophore, to fulfil the electronically pre-resonant Raman condition (Fig. 1c) increasing the Raman-scattering cross-section. This concept was demonstrated experimentally with commercial near-IR absorbing small molecule dyes.[16] The results were striking, with sensitivity limits down to 30–50 molecules in the laser focus for the dye ATTO740, photostable detection, and "super-multiplexing" with 24 resolvable colours. However, to date, there have been no reports expanding this epr-CRS approach to genetically encodable IRFPs. Here, we show that IRFPs, and in particular mRhubarb720, provide strong epr-CRS, and demonstrate their application in cell imaging.

In terms of electronic excitation properties, mRhubarb720 is one of the most red-shifted FPs (absorption peak wavelength $\lambda_{max}$ at 703 nm) with a relatively high molar absorbance ($\varepsilon$=72.5 mM$^{-1}$cm$^{-1}$) compared to other IRFPs such as emIRFP670 and mIRFP670nano3 (Fig. 1c and Table S1). IRFPs are qualitatively different from far-red FPs such as mCherry[7,8] which have classical FP β-barrel structures (Fig. S1) and sequence encoded chromophores that are hypsochromically shifted compared to the IFRPs (Fig. 1b-c).

Using X-ray micro-crystal diffraction data (resolution 3.3 Å) combined with AlphaFold2[19,20] modelling, we determined the structure of mRhubarb720 (Fig. 1d-e and Fig. S2, with Table S2 showing the structural statistics). mRhubarb720 shares a general PAS-GAF two domain structural fold with other biliverdin binding bacterial phytochrome proteins.[21,22] In mRhubarb720, biliverdin is bound to Cys15 in the GAF domain region via a thioether link to the C3 carbon of the A ring (Fig. 1a, 1e and Fig. S2a); the D-ring occupies the 15Za isomeric form. The three main mutations around the chromophore (L196Q, T202D and V203I; see Fig. S2b) are likely to give rise to red-shifted spectral characteristics compared to its predecessor iRFP713.[13] L196Q removes a hydrophobic residue and potentially contributes to a polar side-chain interaction network with E191, K193 and Y198 that plays a role in photoconversion between the two chromophore conformational states, Pr and Pfr [4,23] (Fig. S2c). The T202D mutation results in a negatively charged carboxyl group becoming surface exposed, which is likely to improve solubility. Finally, the V203I mutation extends the interaction between the protein and the C-ring of BV (Fig. S2b).

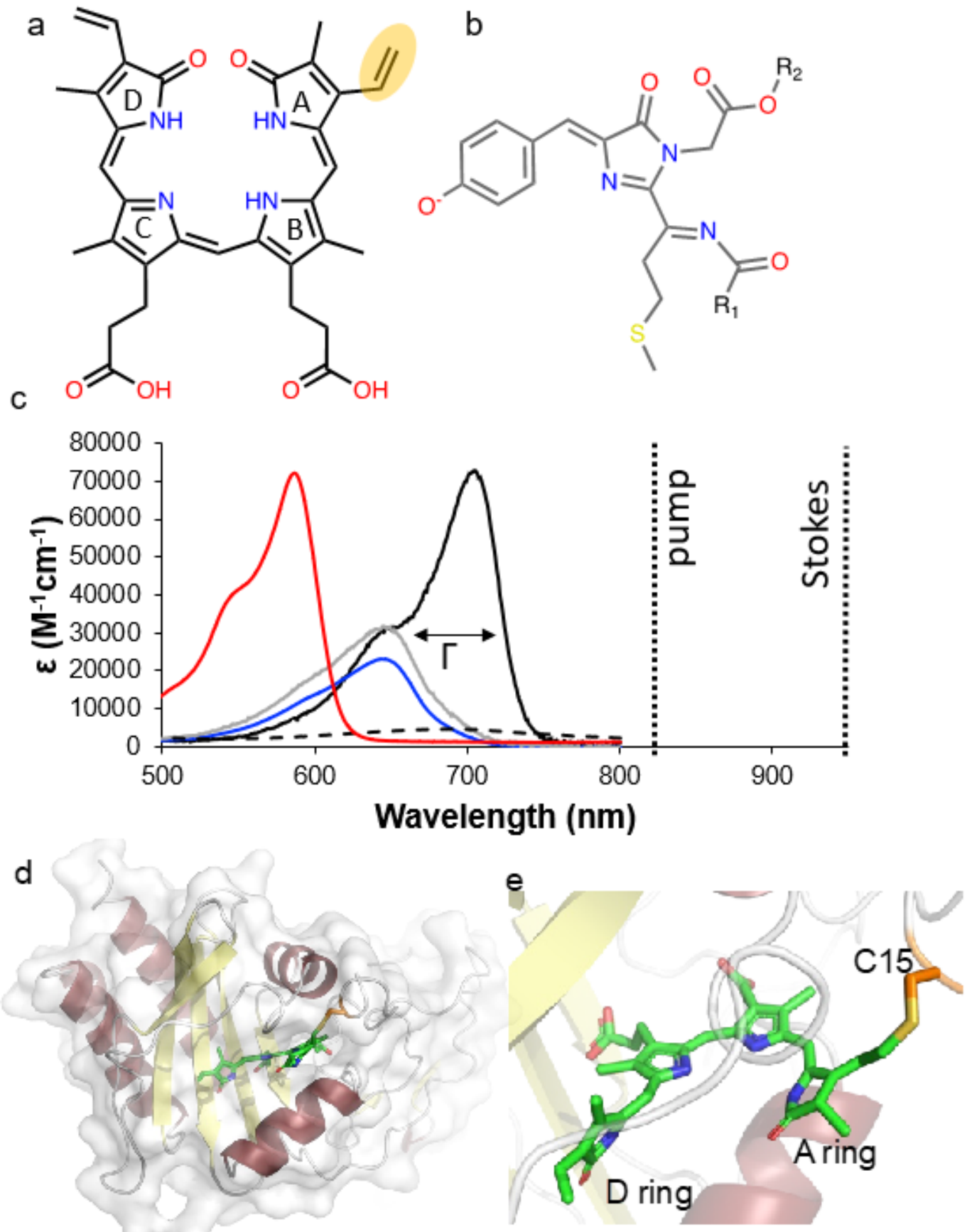

Figure 1. Molecular and spectral features of IRFPs. (a) The biliverdin chromophore common to the investigated IRFPs. The C=C bond highlighted in the orange forms a thioether link to the protein component. (b) The chromophore structure of mCherry. (c) Measured absorbance spectra of mRhubarb720 (black solid), mIRFP670nano3 (grey), emiRFP670 (blue), mCherry (red), and free biliverdin (black dashed). Γ is the linewidth of the electronic transition. The pump and Stokes laser wavelengths in the SRS set-up are shown as dotted lines. (d) The overall molecular structure of mRhubarb determined here, with (e) a close-up of biliverdin (green sticks).

To benefit from electronic pre-resonant CRS, the pump ($\nu_p$) and Stokes ($\nu_S$) laser frequencies used to coherently drive molecular vibrations have to be close to the absorption maximum ($\nu_e$) of the chromophore, while being sufficiently redshifted to suppress photobleaching. The Raman scattering cross-section scales with the frequency detuning[24] as $(\nu_e-\nu_p)^{-4}$. Since $\nu_p$ is closer to the electronic transition frequency than $\nu_S$, the resonance enhancement effect is dominated by the pump. It has been suggested in the literature by Min and colleagues[25] that a suitable excitation frequency is detuned (red-shifted) between 2 and 6 times the homogeneous linewidth Γ of the electronic transition (Fig. 1c). However, it has also been pointed out that the specific properties of the chromophore need to be considered to find an optimal excitation.[26,27] An important consideration is the suppression of one-photon absorption of the pump, and in turn any associated photobleaching. In the low-energy tail of the absorption spectrum, one-photon absorption scales with the probability of finding a molecule in the electronic ground state $S_0$ which is vibrationally excited by thermal occupation. With decreasing photon energy, this probability is exponentially decaying with a slope given by the thermal energy at room temperature of about 205 $cm^{-1}$. For mRhubarb720 at 820 nm pump wavelength, having a detuning of 1790 $cm^{-1}$ from the absorption transition without vibrational excitation, this results in a reduction of the absorption by about 4 orders of magnitude (see also section S7). The Franck-Condon overlap provides a further factor reducing the absorption. This overlap can be estimated from the reduction of the absorption towards the blue side of the absorption peak in Fig. 1c, which is dominated by the creation of vibrations in the electronic excited state $S_1$ and does not require thermal occupation. A reduction by a factor around 3 for the given pump detuning can be deduced.

In addition to one-photon absorption, two-photon absorption (TPA) of the exciting pump and Stokes beams has to be considered as an electronic excitation mechanism that may lead to photobleaching. TPA is also a source of background in CRS. For example, in SRS measured as stimulated Raman loss (SRL) in the pump beam transmission, pump-Stokes TPA manifests as an apparent signal, due to the additional pump photon loss mediated by this absorption. Notably, TPA is, as CRS, a third-order non-linear optical process. Therefore, it scales similar to CRS with the laser excitation pulse duration and power and thus cannot be supressed relative to CRS using these parameters. Typical TPA cross-section spectra are broad, reflecting the properties of the higher electronic states of the molecules, and TPA accordingly provides a broad background in the CRS spectra. Notably, the best-performing epr-CRS probes in terms of signal to background ratio and photostability reported so far are close to inversion symmetric, which is supressing TPA, as discussed in section S7 in the Supporting Information.

The SRS setup used in this study is shown in Fig. S3 and operates with a fixed pump wavelength centred at 820 nm and a tuneable Stokes wavelength, adjusted according to the vibrational resonance range of interest, for example at 956 nm (Fig. 1c) resulting in a centre wavenumber of 1740 $cm^{-1}$. This pump wavelength is fulfilling the epr-CRS condition for mRhubarb720 with an absorption peak wavelength $\lambda_{max}$ of 703 nm. For the mIRFP670nano3, emiRFP670 and mCherry chromophores, $\nu_e$ is blue-shifted, hence we expect a lower epr-CRS effect (see Fig. S4). To measure SRL spectra we use spectral focussing as detailed previously[28] and outlined in section S5 of the Supporting Information. In this approach, pump and Stokes pulses of about 100 fs duration are equally linearly chirped to about 3 ps duration, providing a spectral resolution of 11 $cm^{-1}$. This resolution is optimal for the resonances probed as discussed in section S5 (also see Fig. S5).

By changing the delay time between the pulses, the instantaneous frequency difference (IFD) is tuned, allowing to record an SRL spectrum, as shown in Fig. 2a (dashed line) for mRhubarb720. As mentioned, when pump and Stokes pulses are in temporal overlap, a combined two-photon absorption process occurs (namely the absorption of $\nu_p+\nu_S$, see Fig. S6 and S7) which manifests as a pump-loss and thus appears as SRL signal in our configuration (we detect the stimulated Raman loss in the pump and intensity modulate the Stokes; Fig. S3). This TPA pump-Stokes cross-correlation profile has been subtracted from the data, to highlight the vibrational SRL contribution, and corresponding spectra are shown in Fig. 2 as solid lines (see Fig. S8 and S9 for all measured spectra before subtraction and section S5 for procedural details). Vibrational resonances around 1620-1650 $cm^{-1}$ are observed, characteristic of the C=C and C=N bonds which dominate the biliverdin chromophore (Fig. 1a). Notably, when tuning the Stokes wavelength to a centre wavenumber of 1640 $cm^{-1}$ (blue line Fig.2a), these resonances are more efficiently driven when compared to using 1740 $cm^{-1}$ (black line Fig.2a). This is because the centre wavenumber corresponds to the maximum temporal overlap between pump and Stokes (see also Fig. S8 and S10). Hence, when the centre is tuned near the vibrational resonance of interest, an optimal use of the excitation intensity pulse overlap is obtained. Indeed, considering the spectra for mRhubarb720 in Fig.2a, we see a 7-fold increase in the SRL signal for the 1620 $cm^{-1}$ resonance when using 1640 $cm^{-1}$ as the centre IFD compared to 1740 $cm^{-1}$. We verified that the SRL signal scales linearly with the pump power, and that the signal disappears when blocking one of the beams, as expected (Fig. S6). For comparison, we measured the SRL spectrum of free BV in solution (Fig. 2a), i.e. not bound to an IRFP. Interestingly, we found a spectrum lacking any significant vibrational resonance. We attribute this difference to the solvation environment of free BV, as compared to BV bound to the protein component, resulting in a much weaker, broader and shifted absorption spectrum (see Fig.1c and S11a), supressing the epr enhancement of SRS. The comparison of the SRL spectra in Fig. S11b shows a hundredfold reduction of the signal from free BV and a vibrational response broadened and shifted to 1550-1620 $cm^{-1}$. Simply put, binding of BV to the protein provides a stable environment and maintains BV in a defined conformation, allowing for enhanced absorption and longer vibrational coherence. Importantly, the lack of significant SRL from free BV enables detecting IRFPs with negligible SRL background from any unbound cofactor naturally present in cells.

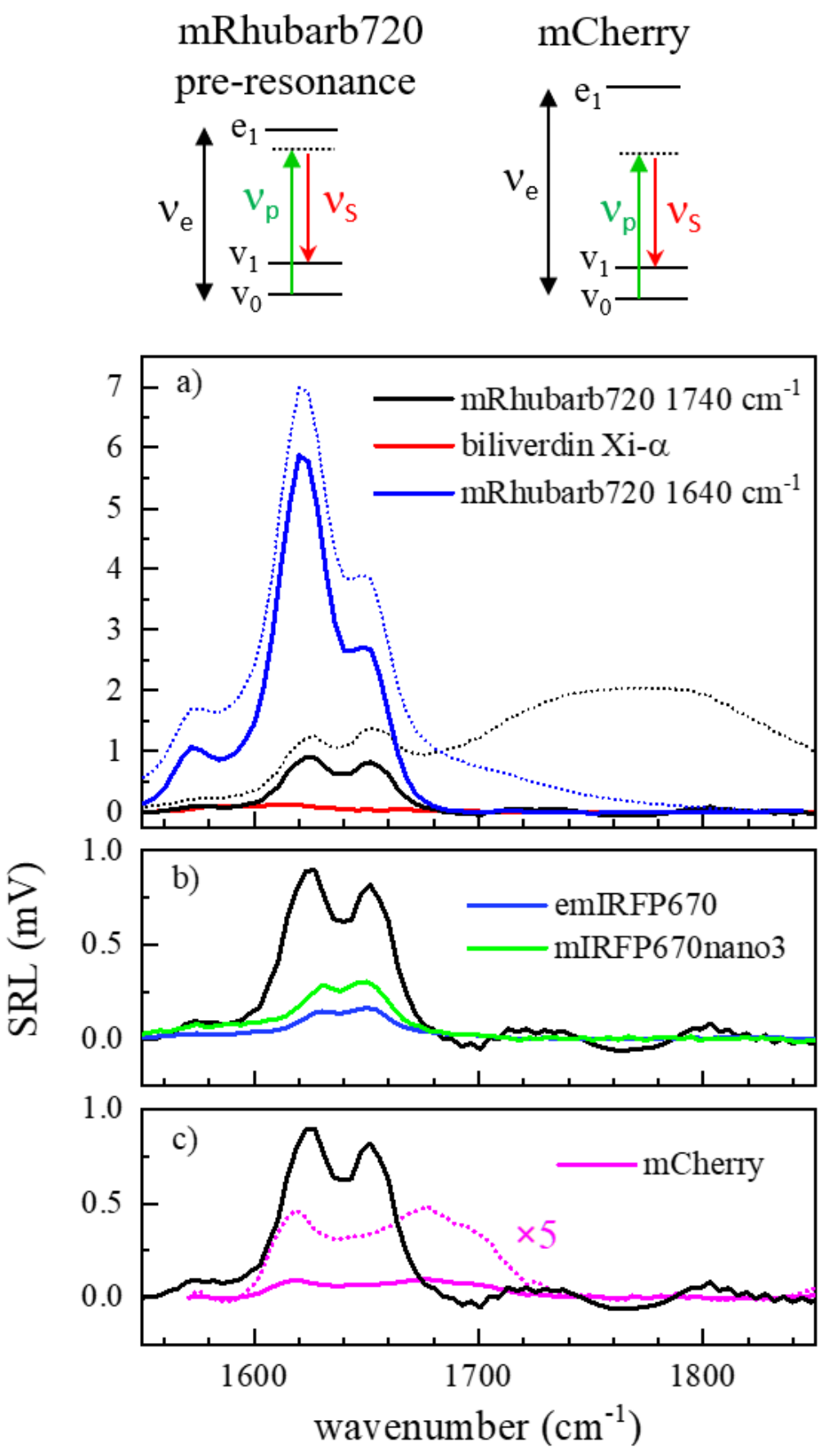

Figure 2: Stimulated Raman loss spectra of IRFPs in solution (1 mM concentration in 50 mM Tris buffer, 1740 $cm^{-1}$ centre IFD). (a) BV with 50 mM NaOH (red), mRhubarb720 (black), also at 1640 $cm^{-1}$ centre IFD (blue); dotted lines as measured, solid lines TPA subtracted. (b) BV-containing IRFPs mRhubarb720 (black), mIRFP670nano3 (green) and emIRFP670 (blue). (c) mCherry (magenta, dotted line scaled by factor 5) compared to mRhubarb720 (black). Top: diagrams illustrating the energy separation between the electronic transition and the pump frequency in our SRL set-up, for mRhubarb720 and mCherry. $\nu_e$ is the electronically resonant absorption frequency of the chromophore, $v_0$ and $v_1$ represent the ground and excited vibrational states, respectively, whereas $e_1$ is the first excited electronic state. Microscope objective 20× 0.75 NA, pump (Stokes) power at the sample 6 mW (8 mW), 0.1 ms pixel dwell time, averaged to 3 s per spectral point.

The results in Fig.2 show that mRhubarb720 exhibits the largest epr-SRS effect, compared to emIRFP670, mIRFP670nano3, and mCherry, as expected from the resonant absorption wavelength being closer to the pump wavelength. Note that emIRFP670 and mIRFP670nano3 also use BV as chromophore but have an absorption peak $\lambda_{max}$ of 644 and 645 nm, respectively, hence blue-shifted compared to mRhubarb720 at 703 nm, with a frequency difference $\nu_e$-$\nu_p$ of about 3300 $cm^{-1}$ compared to 2030 $cm^{-1}$ for mRhubarb720 (Fig. 2b). While similar bond types are present in mCherry's chromophore, it has a different chromophore (Fig. 1b) and a $\lambda_{max}$ at about 590 nm (Fig. 1c), so is further away from the pump wavelength in the SRL experiment, corresponding to a frequency difference of 4754 $cm^{-1}$. As a result, the SRL peak signal of mRhubarb720 is about 10-fold higher than for mCherry (Fig. 2c).

The epr-SRL signal is expected to scale as the Raman cross-section. For a single dominant transition, the frequency dependent factor is given[29] by $F_A = \left(\nu_p - \nu_v\right)^2 \left(\nu_e^2 + \nu_p^2\right) \left(\nu_e^2 - \nu_p^2\right)^{-2}$ and is shown in Fig. S4 for the experimental parameters used here. The Raman cross-section and thus the SRL signal scales with $F_A^2$ , and compared to mRhubarb720 a 7-fold reduction for IRFP670 and a 29-fold reduction for mCherry is expected. These factors have the same trend but are somewhat larger than the experimentally measured ones. The frequency dependent factor was explored for a range of fluorophores and pump wavelengths in the literature[30,31], also showing a somewhat weaker dependence than expected, and variations between fluorophores. With increasing frequency, an increase of the coupling to the vibration expressed as vibronic energy (Eqn. 4b in [32]) might be expected, due to a reduced extension of the electronic excitation over the conjugated chains, counteracting the decreasing resonance enhancement and explaining the observed behaviour.

Considering the excitation and detection conditions in our experiment, we have estimated the detection sensitivity limit for mRhubarb720 (see also Section S9). The SRL relative intensity modulation is found to be $\Delta I/I = 2.6\times10^{-5}$ at 1 mM protein concentration. This is similar to the value reported for the ATTO740 dye previously[33], when scaled to the laser power and concentration conditions used here. The photon shot-noise limit in our experiment (as obtained from the measured DC current) is found to be $1.2\times10^{-8}/\sqrt{\text{Hz}}$. At 500Hz bandwidth (corresponding to 1 ms per pixel) this equates to 1.9×10$^{-7}$ (for one quadrature of the detection) which results in a detection limit of 7.2 µM concentration. When further scaling to the powers used by Wei and Min[33] we find a shot-noise limited detection equating to 452 nM. This is comparable to the detection limit for ATTO740 of 250 nM reported previously.[33] Hence, mRhubarb720 appears to be an SRS tag with a signal strength comparable to the infrared dye ATTO740 characterised previously,[33] while bringing the much sought after genetic encoding capability. We then used mRhubarb720 for cell imaging, to demonstrate its application as genetically encoded Raman tag.

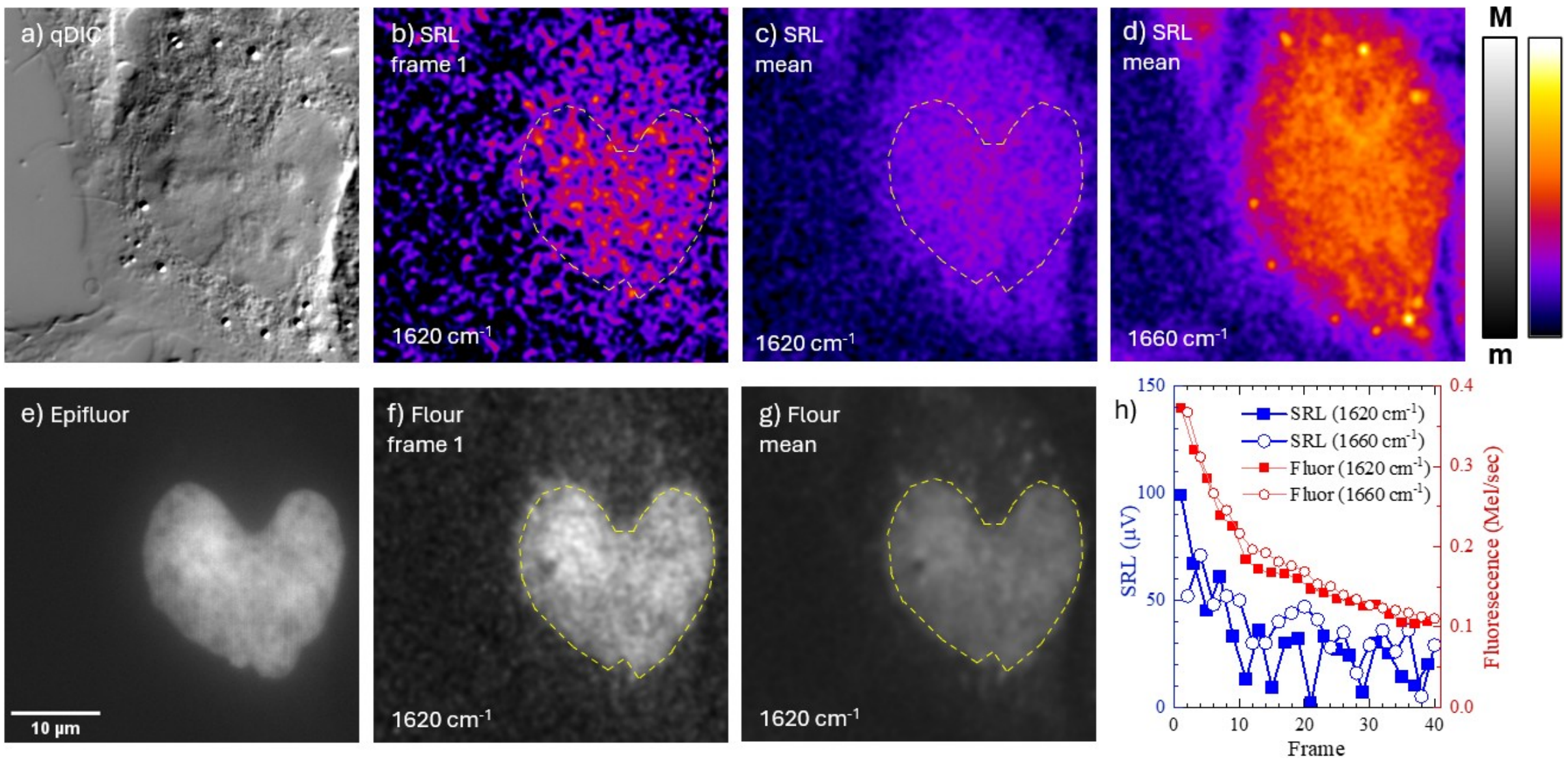

Figure 3: Imaging HeLa cell H2B-labelled mRhubarb720 expression. (a) qDIC, (b-d) SRL, (e) widefield epi-fluorescence (Epifluor) and (f-g) forward detected fluorescence (Fluor) are shown, as indicated. Grey and colour scales are linear from m to M as follows: qDIC (m = 0.7 rad to M = 0.7 rad), SRL (m = 0.9 mV to M = 2.2 mV), Epifluor (m = 0 pe to M = 4 kpe, where pe are photoelectrons), Fluor (m = 0 Mpe/s to M = 1 Mpe/s). SRL and Fluor are taken over 40 frames, alternating between 1620 and 1660 cm$^{-1}$, 0.01 ms pixel dwell time, 50×50 µm at 463×463 points, about 2s per frame. qDIC and Epifluor is taken before CRS imaging. The mean is taken over the 20 frames using the wavenumber given. Microscope objective 60× 1.27 NA, pump (Stokes) power at the sample 7 mW (12 mW). Centre IFD 1640 cm$^{-1}$. (h) the spatially averaged SRL and Fluor over the heart-shaped nucleus, as function of frame number, separated for the two wavenumbers. The SRL has been background-subtracted using the non-fluorescent cell regions (see Fig. S12).

We employed mRhubarb720 as a classical fusion tag to facilitate imaging in mammalian cells. In this case, mRhubarb720 was fused to the C-terminal of the histone protein H2B, a structural protein of eukaryotic chromosomes. In HeLa cells (Fig. 3a), quantitative differential interference contrast (qDIC) microscopy (for imaging methods see section S5) shows the cell structure with a "heart shaped" nucleus, nucleoli, and the cytosol containing small lipid droplets. Epi-fluorescence (Epifluor) microscopy (Fig. 3e) highlights the nucleus as expected due to the localisation of H2B-mRhubarb720. SRL (Fig. 3b-d) and forward fluorescence (Fluor; Fig. 3f-g) were then measured simultaneously under pulsed laser excitation, using galvo beam scanning for imaging, and a photomultiplier with appropriate bandpass filters for fluorescence (see section S5). SRL measurements used 1640cm$^{-1}$ as centre IFD to optimally drive vibrational resonances. They were alternating between IFD wavenumbers of 1620 cm$^{-1}$, the peak of the mRubarb720 SRL, and 1660 cm$^{-1}$ where there is weaker mRhubarb SRL (see Fig.2a blue lines), but significant SRL from endogenous cellular components, e.g. C=C bonds in lipids.[34]

Forward fluorescence detection for mRhubarb720 in solution (Fig. S6) suggests that this fluorescence is dominated by one-photon absorption of the pump-only beam, which is possible due to the pre-resonant condition. As mentioned, albeit the pump beam has an energy below the absorption peak of mRhubarb720, at room temperature the thermal population of higher energy vibrational states enables one-photon absorption of the pump into the first electronic excited state, and subsequent fluorescence (see section S7 of the Supporting Information). Upon imaging, forward fluorescence accordingly shows the nucleus geometry, similar to wide-field epi-fluorescence. There is significant fluorescence bleaching during the time-course acquisition repeated over several frames (see red lines in Fig. 3h), indicating that the one and two-photon absorption of mRhubarb720 creates real excitations leading to photobleaching.

The first SRL frame acquired at 1620 cm$^{-1}$ shows a more specific nucleus contrast, though with low signal to noise (Fig. 3b). The mean SRL signal at 1620cm$^{-1}$ averaged over all acquired frames is however weak and shows a cell outline where the nucleus cannot be clearly distinguished (Fig 3c). At 1660cm$^{-1}$ the cell outline is more visible, and lipid droplets appear (Fig. 3d). The SRL signal at 1620 cm$^{-1}$ exhibits a similar photobleaching behaviour as fluorescence (see the blue filled squares in Fig.3h). For SRL acquired at 1660 cm$^{-1}$ this bleaching is slightly reduced but still present (blue circles in Fig. 3h). The observed SRL signal at 1620 cm$^{-1}$ is about 0.08mV in the first frame, corresponding to a concentration of mRhubarb720 of about 10uM, estimated from the solution measurements (Fig. 2). Similar results are shown in Fig. S13 on another HeLa cell example. Notably, while the demonstration of epr-SRS in solution is promising (as shown in Fig. 2), in this scenario proteins are free to diffuse in and out the focal volume and refill any bleached product. Conversely, when fixed inside cells, mRhubarb720 photobleaching upon repeated exposure limits its application in fluorescence and in SRS imaging. Therefore, future efforts will have to address FP designs with reduced photobleaching and hence more robust upon epr-CRS excitation conditions.

In conclusion, we have demonstrated the first steps to genetically encoding probes for use in cellular based electronically pre-resonance coherent Raman imaging. The relatively narrow spectral linewidths of these probes, coupled with the electronic pre-resonant effect, exhibiting signal enhancements comparable to what shown in the literature for synthetic ATTO dyes, open a promising avenue towards super-multiplexing CRS with genetic encoding. As the basis for the probes are existing fluorescent proteins, the transition of FPs to SRS imaging is relatively straightforward from a cell biology point of view. The next step is to further develop these probes as well as microscopy setups. Microscopy developments include SRS set-ups featuring pump wavelengths in the visible, so that FPs with chromophores absorbing in the blue to red range can be used, as well as SRS multiplex acquisition, for detection of multiple probes. Furthermore, SRS implementations using lower repetition rate lasers and faster beam scanning would enable reduction of photobleaching by applying only a few exciting pulses to any given position. Probe developments include reducing two-photon absorption cross-sections, to suppress spectrally broad background and photobleaching, and further expanding the utility of such probes via the incorporation of electronically-coupled Raman bonds vibrating in the biologically silent region, e.g. via reprogrammed genetic code approaches.[35–38]

## ASSOCIATED CONTENT

### Supporting Information

Supporting information is available free of charge on the ACS Publications website. It contains detailed methods, gene sequences, Supporting Tables S1-S2 and Supporting Figures S1-S15.

## AUTHOR INFORMATION

### Corresponding Authors

* Paola Borri. School of Biosciences, Sir Martin Evans Building, Cardiff University, Cardiff, UK. CF10 3AX. Email: borrip@cardiff.ac.uk. Tel: +44 2920879356
* D. Dafydd Jones. School of Biosciences, Sir Martin Evans Building, Cardiff University, Cardiff, UK. CF10 3AX. Email: jonesdd@cardiff.ac.uk. Tel: +44 2920874290.
* Wolfgang Langbein. School of Physics and Astronomy, Cardiff University, Cardiff, UK. CF24 3AA. Email: langbeinww@cardiff.ac.uk. Tel: +44 2920870172

### Present Addresses

†If an author's address is different than the one given in the affiliation line, this information may be included here.

### Author Contributions

The manuscript was written through contributions of all authors.
All authors have given approval to the final version of the manuscript. ‡These authors contributed equally.

### Funding Sources

The authors acknowledge funding from the EPSRC (EP/V048147/1, EP/M028313/1), BBSRC (BB/Z514913/1). O.Z. was supported by an EPSRC DTP studentship.

### Data availability

Information on the data underpinning the results presented here, including how to access them, are shared via FigShare (https://figshare.com/s/3503c0259bd34acd54e3).

## ACKNOWLEDGMENT

We would like to thank the staff at the Diamond Light Source (Harwell, UK) for the supply of facilities and beam time, especially microfocus macromolecular crystallography (MX) beamline (VMXm). We would also like to thank the Protein Technology Hub facility in the School of Biosciences, Cardiff University for access to protein purification and analysis facilities, and the Cardiff University Bioimaging Hub Core Facility, RRID:SCR_022556. We thank Dr Iestyn Pope for contributions to the CRS microscope setups. We would like to thank Prof Andreas Zumbusch, University of Konstanz for insightful discussions regarding epr-SRS using fluorescent proteins.

## ABBREVIATIONS

BV, biliverdin Xi-α; qDIC, quantitative Differential interference contrast; FP, fluorescent protein; epr-CRS, electronically pre-resonant coherent Raman scattering; IRFP, infrared fluorescent protein; SRS, stimulated Raman scattering; SRL stimulated Raman loss, $\nu_P$, pump laser frequency; $\nu_S$, Stokes laser frequency; $\nu_e$ chromophore electronic absorption peak frequency.

## REFERENCES

(1) Shaner, N. C.; Steinbach, P. A.; Tsien, R. Y. A Guide to Choosing Fluorescent Proteins. *Nat Methods* 2005, *2* (12), 905–909. https://doi.org/10.1038/nmeth819.

(2) Rodriguez, E. A.; Campbell, R. E.; Lin, J. Y.; Lin, M. Z.; Miyawaki, A.; Palmer, A. E.; Shu, X.; Zhang, J.; Tsien, R. Y. The Growing and Glowing Toolbox of Fluorescent and Photoactive Proteins. *Trends Biochem Sci* 2017, *42* (2), 111–129. https://doi.org/10.1016/j.tibs.2016.09.010.

(3) Tsien, R. Y. The Green Fluorescent Protein. *Annu Rev Biochem* 1998, *67*, 509–544. https://doi.org/10.1146/annurev.biochem.67.1.509.

(4) Rogers, O. C.; Johnson, D. M.; Firnberg, E. MRhubarb: Engineering of Monomeric, Red-Shifted, and Brighter Variants of IRFP Using Structure-Guided Multi-Site Mutagenesis. *Sci Rep* 2019, *9* (1). https://doi.org/10.1038/s41598-019-52123-7.

(5) Matlashov, M. E.; Shcherbakova, D. M.; Alvelid, J.; Baloban, M.; Pennacchietti, F.; Shemetov, A. A.; Testa, I.; Verkhusha, V. V. A Set of Monomeric Near-Infrared Fluorescent Proteins for Multicolor Imaging across Scales. *Nat Commun* 2020, *11* (1). https://doi.org/10.1038/s41467-019-13897-6.

(6) Oliinyk, O. S.; Baloban, M.; Clark, C. L.; Carey, E.; Pletnev, S.; Nimmerjahn, A.; Verkhusha, V. V. Single-Domain near-Infrared Protein Provides a Scaffold for Antigen-Dependent Fluorescent Nanobodies. *Nat Methods* 2022, *19* (6), 740–750. https://doi.org/10.1038/s41592-022-01467-6.

(7) Shu, X.; Shaner, N. C.; Yarbrough, C. A.; Tsien, R. Y.; Remington, S. J. Novel Chromophores and Buried Charges Control Color in MFruits. *Biochemistry* 2006, *45* (32), 9639–9647. https://doi.org/10.1021/bi060773l.

(8) Shaner, N. C.; Campbell, R. E.; Steinbach, P. A.; Giepmans, B. N. G.; Palmer, A. E.; Tsien, R. Y. Improved Monomeric Red, Orange and Yellow Fluorescent Proteins Derived from Discosoma Sp. Red Fluorescent Protein. *Nat Biotechnol* 2004, *22* (12), 1567–1572. https://doi.org/10.1038/nbt1037.

(9) Zhang, Y.; Ling, J.; Liu, T.; Chen, Z. Lumos Maxima – How Robust Fluorophores Resist Photobleaching? *Curr Opin Chem Biol* 2024, *79*, 102439. https://doi.org/10.1016/j.cbpa.2024.102439.

(10) Ganini, D.; Leinisch, F.; Kumar, A.; Jiang, J.; Tokar, E.; Malone, C. C.; Petrovich, R. M.; Mason, R. P. Fluorescent Proteins Such as EGFP Catalytically Generate Superoxide Anion Free Radical and H2O2 in the Presence of NAD(P)H. *Free Radic Biol Med* 2016, *100*, S23. https://doi.org/10.1016/j.freeradbiomed.2016.10.055.

(11) Trewin, A. J.; Berry, B. J.; Wei, A. Y.; Bahr, L. L.; Foster, T. H.; Wojtovich, A. P. Light-Induced Oxidant Production by Fluorescent Proteins. *Free Radic Biol Med* 2018, *128*, 157–164. https://doi.org/10.1016/j.freeradbiomed.2018.02.002.

(12) Shaner, N. C.; Lin, M. Z.; McKeown, M. R.; Steinbach, P. A.; Hazelwood, K. L.; Davidson, M. W.; Tsien, R. Y. Improving the Photostability of Bright Monomeric Orange and Red Fluorescent Proteins. *Nat Methods* 2008, *5* (6), 545–551. https://doi.org/10.1038/nmeth.1209.

(13) Filonov, G. S.; Piatkevich, K. D.; Ting, L.-M.; Zhang, J.; Kim, K.; Verkhusha, V. V. Bright and Stable Near-Infrared Fluorescent Protein for in Vivo Imaging. *Nat Biotechnol* 2011, *29* (8), 757–761. https://doi.org/10.1038/nbt.1918.

(14) Camp, C. H.; Cicerone, M. T. Chemically Sensitive Bioimaging with Coherent Raman Scattering. *Nature Photonics*. Nature Publishing Group May 30, 2015, pp 295–305. https://doi.org/10.1038/nphoton.2015.60.

(15) Efremov, E. V.; Ariese, F.; Gooijer, C. Achievements in Resonance Raman Spectroscopy. *Anal Chim Acta* 2008, *606* (2), 119–134. https://doi.org/10.1016/j.aca.2007.11.006.

(16) Wei, L.; Chen, Z.; Shi, L.; Long, R.; Anzalone, A. V; Zhang, L.; Hu, F.; Yuste, R.; Cornish, V. W.; Min, W. Super-Multiplex Vibrational Imaging. *Nature* 2017, *544* (7651), 465–470. https://doi.org/10.1038/nature22051.

(17) Pruccoli, A.; Kocademir, M.; Winterhalder, M. J.; Zumbusch, A. Electronically Preresonant Stimulated Raman Scattering Microscopy of Weakly Fluorescing Chromophores. *J Phys Chem B* 2023, *127* (27), 6029–6037. https://doi.org/10.1021/acs.jpcb.3c01407.

(18) Wei, L.; Min, W. Electronic Preresonance Stimulated Raman Scattering Microscopy. *J Phys Chem Lett* 2018, *9* (15), 4294–4301. https://doi.org/10.1021/acs.jpclett.8b00204.

(19) Jumper, J.; Evans, R.; Pritzel, A.; Green, T.; Figurnov, M.; Ronneberger, O.; Tunyasuvunakool, K.; Bates, R.; Žídek, A.; Potapenko, A.; Bridgland, A.; Meyer, C.; Kohl, S. A. A.; Ballard, A. J.; Cowie, A.; Romera-Paredes, B.; Nikolov, S.; Jain, R.; Adler, J.; Back, T.; Petersen, S.; Reiman, D.; Clancy, E.; Zielinski, M.; Steinegger, M.; Pacholska, M.; Berghammer, T.; Bodenstein, S.; Silver, D.; Vinyals, O.; Senior, A. W.; Kavukcuoglu, K.; Kohli, P.; Hassabis, D. Highly Accurate Protein Structure Prediction with AlphaFold. *Nature 2021 596:7873* 2021, *596* (7873), 583–589. https://doi.org/10.1038/s41586-021-03819-2.

(20) Mirdita, M.; Schütze, K.; Moriwaki, Y.; Heo, L.; Ovchinnikov, S.; Steinegger, M. ColabFold: Making Protein Folding Accessible to All. *Nat Methods* 2022, *19* (6), 679–682. https://doi.org/10.1038/s41592-022-01488-1.

(21) Wagner, J. R.; Brunzelle, J. S.; Forest, K. T.; Vierstra, R. D. A Light-Sensing Knot Revealed by the Structure of the Chromophore-Binding Domain of Phytochrome. *Nature* 2005, *438* (7066), 325–331. https://doi.org/10.1038/nature04118.

(22) Baloban, M.; Shcherbakova, D. M.; Pletnev, S.; Pletnev, V. Z.; Lagarias, J. C.; Verkhusha, V. V. Designing Brighter Near-Infrared Fluorescent Proteins: Insights from Structural and Biochemical Studies. *Chem Sci* 2017, *8* (6), 4546–4557. https://doi.org/10.1039/C7SC00855D.

(23) Yang, X.; Kuk, J.; Moffat, K. Conformational Differences between the Pfr and Pr States in *Pseudomonas Aeruginosa* Bacteriophytochrome. *Proceedings of the National Academy of Sciences* 2009, *106* (37), 15639–15644. https://doi.org/10.1073/pnas.0902178106.

(24) Pruccoli, A.; Zumbusch, A. High Sensitivity Stimulated Raman Scattering Microscopy with Electronic Resonance Enhancement. *ChemPhysChem* 2024, *25* (19). https://doi.org/10.1002/cphc.202400309.

(25) Wei, L.; Chen, Z.; Shi, L.; Long, R.; Anzalone, A. V.; Zhang, L.; Hu, F.; Yuste, R.; Cornish, V. W.; Min, W. Super-Multiplex Vibrational Imaging. *Nature* 2017, *544* (7651), 465–470. https://doi.org/10.1038/nature22051.

(26) Pruccoli, A.; Zumbusch, A. High Sensitivity Stimulated Raman Scattering Microscopy with Electronic Resonance Enhancement. *ChemPhysChem* 2024, *25* (19). https://doi.org/10.1002/cphc.202400309.

(27) Pruccoli, A.; Kocademir, M.; Winterhalder, M. J.; Zumbusch, A. Electronically Preresonant Stimulated Raman Scattering Microscopy of Weakly Fluorescing Chromophores. *J Phys Chem B* 2023, *127* (27), 6029–6037. https://doi.org/10.1021/acs.jpcb.3c01407.

(28) Langbein, W.; Regan, D.; Pope, I.; Borri, P. Invited Article: Heterodyne Dual-Polarization Epi-Detected CARS Microscopy for Chemical and Topographic Imaging of Interfaces. *APL Photonics* 2018, *3* (9). https://doi.org/10.1063/1.5027256.

(29) Albrecht, A. C.; Hutley, M. C. On the Dependence of Vibrational Raman Intensity on the Wavelength of Incident Light. *J Chem Phys* 1971, *55* (9), 4438–4443. https://doi.org/10.1063/1.1676771.

(30) Wei, L.; Chen, Z.; Shi, L.; Long, R.; Anzalone, A. V.; Zhang, L.; Hu, F.; Yuste, R.; Cornish, V. W.; Min, W. Super-Multiplex Vibrational Imaging. *Nature* 2017, *544* (7651), 465–470. https://doi.org/10.1038/nature22051.

(31) Pruccoli, A.; Zumbusch, A. High Sensitivity Stimulated Raman Scattering Microscopy with Electronic Resonance Enhancement. *ChemPhysChem* 2024, *25* (19). https://doi.org/10.1002/cphc.202400309.

(32) Albrecht, A. C.; Hutley, M. C. On the Dependence of Vibrational Raman Intensity on the Wavelength of Incident Light. *J Chem Phys* 1971, *55* (9), 4438–4443. https://doi.org/10.1063/1.1676771.

(33) Wei, L.; Min, W. Electronic Preresonance Stimulated Raman Scattering Microscopy. *J Phys Chem Lett* 2018, *9* (15), 4294–4301. https://doi.org/10.1021/acs.jpclett.8b00204.

(34) Di Napoli, C.; Pope, I.; Masia, F.; Langbein, W.; Watson, P.; Borri, P. Quantitative Spatiotemporal Chemical Profiling of Individual Lipid Droplets by Hyperspectral CARS Microscopy in Living Human Adipose-Derived Stem Cells. *Anal Chem* 2016, *88* (7), 3677–3685. https://doi.org/10.1021/acs.analchem.5b04468.

(35) Reddington, S. C.; Baldwin, A. J.; Thompson, R.; Brancale, A.; Tippmann, E. M.; Jones, D. D. Directed Evolution of GFP with Non-Natural Amino Acids Identifies Residues for Augmenting and Photoswitching Fluorescence. *Chem Sci* 2015, *6* (2), 1159–1166. https://doi.org/10.1039/c4sc02827a.

(36) Reddington, S. C.; Rizkallah, P. J.; Watson, P. D.; Pearson, R.; Tippmann, E. M.; Jones, D. D. Different Photochemical Events of a Genetically Encoded Phenyl Azide Define and Modulate GFP Fluorescence. *Angewandte Chemie - International Edition* 2013, *52* (23), 5974–5977. https://doi.org/10.1002/anie.201301490.

(37) Reddington, S. C.; Driezis, S.; Hartley, A. M.; Watson, P. D.; Rizkallah, P. J.; Jones, D. D. Genetically Encoded Phenyl Azide Photochemistry Drives Positive and Negative Functional Modulation of a Red Fluorescent Protein. *RSC Adv* 2015, *5* (95), 77734–77738. https://doi.org/10.1039/c5ra13552d.

(38) Romei, M. G.; Lin, C. Y.; Mathews, I. I.; Boxer, S. G. Electrostatic Control of Photoisomerization Pathways in Proteins. *Science (1979)* 2020, *367* (6473), 76–79. https://doi.org/10.1126/science.aax1898.

# Genetically encoding stimulated Raman-scattering probes for cell imaging using infrared fluorescent proteins – Supporting information

David Regan[1]‡, Ozan Aksakal[1]‡, Athena Zitti[1], John McLarnon[1], Magdalena Lipka-Lloyd[2], Pierre J. Rizkallah[2], Anna J. Warren[3], Peter D. Watson[1], Wolfgang Langbein[4*], D. Dafydd Jones[1*], Paola Borri[1*]

1. Molecular Bioscience Division, School of Biosciences, Cardiff University, Cardiff, UK. 2. School of Medicine, Cardiff University, Cardiff, UK. 3. Diamond Light Source Ltd, Harwell Science and Innovation Campus, Harwell, UK. 4. School of Physics and Astronomy, Cardiff University, Cardiff, UK.

## Section S1 Gene sequences:

### emiRFP670

*Near-IR FP*

ATGGGGAGCCACCATCACCATCACCATGGCAGATCTATGGCGGAAGGCTCCGTCGCCAGGCAGCCTG
ACCTCTTGACCTGCGAACATGAAGAGATCCACCTCGCCGGCTCGATCCAGCCGCATGGCGCGCTTCTG
GTCGTCAGCGAACATGATCATCGCGTCATCCAGGCCAGCGCCAACGCCGCGGAATTTCTGAATCTCGG
AAGCGTACTCGGCGTTCCGCTCGCCGAGATCGACGGCGATCTGTTGATCAAGATCCTGCCGCATCTCG
ATCCCACCGCCGAAGGCATGCCGGTCGCGGTGCGCTGCCGGATCGGCAATCCCTCTACGGAGTACTGC
GGTCTGATGCATCGGCCTCCGGAAGGCGGGCTGATCATCGAACTCGAACGTGCCGGCCCGTCGATCG
ATCTGTCAGGCACGCTGGCGCCGGCGCTGGAGCGGATCCGCACGGCGGGTTCACTGCGCGCGCTGT
GCGATGACACCGTGCTGCTGTTTCAGCAGTGCACCGGCTACGACCGGGTGATGGTGTATCGTTTCGAT
GAGCAAGGCCACGGCCTGGTATTCTCCGAGTGCCATGTGCCTGGGCTCGAATCCTATTTCGGCAACCG
CTATCCGTCGTCGACTGTCCCGCAGATGGCGCGGCAGCTGTACGTGCGGCAGCGCGTCCGCGTGCTG
GTCGACGTCACCTATCAGCCGGTGCCGCTGGAGCCGCGGCTGTCGCCGCTGACCGGGCGCGATCTCG
ACATGTCGGGCTGCTTCCTGCGCTCGATGTCGCCGTGCCATCTGCAGTTCCTGAAGGACATGGGCGTG
CGCGCCACCCTGGCGGTGTCGCTGGTGGTCGGCGGCAAGCTGTGGGGCCTGGTTGTCTGTCACCATT
ATCTGCCGCGCTTCATCCGTTTCGAGCTGCGGGCGATCTGCAAACGGCTCGCCGAAAGGATCGCGAC
GCGGATCACCGCGCTTGAGAGCTAA

*DNA (T7 promoter for HO-1)*

GAGCTCGGCGCGCCTGCAGGTCGACAAGCTTGCGGCCGCATAATGCTTAAGTCGAACAGAAAGTAAT
CGTATTGTACACGGCCGCATAATCGAAATTAATACGACTCACTATAGGGGAATTGTGAGCGGATAACAA
TTCCCCATCTTAGTATATTAGTTAAGTATAAGAAGGAGATATACAT

*HO-1*

ATGACGAATCTTGCGCAGAAACTCAGGTATGGAACCCAACAGAGCCATACCTTGGCAGAGAACACCG
CATACATGAAGTGCTTCCTTAAAGGCATTGTTGAACGTGAACCTTTTAGACAGCTGCTCGCCAATTTGT
ACTACCTTTATTCCGCACTGGAGGCCGCACTCCGGCAGCATCGCGACAATGAGATCATCAGTGCAATCT
ATTTTCCTGAACTGAATCGTACGGACAAGCTCGCAGAGGATCTGACGTATTATTACGGTCCGAATTGGC
AGCAAATCATTCAGCCAACACCGTGCGCCAAAATATATGTAGATCGATTGAAAACGATTGCTGCATCGG
AGCCGGAATTGCTTATCGCTCACTGTTATACCCGCTATCTTGGTGATTTGTCGGGCGGCCAAAGTCTTA
AAAACATTATTCGTTCAGCACTGCAGTTGCCCGAAGGTGAAGGGACGGCTATGTATGAGTTCGACAGC

CTGCCAACACCGGGAGATCGCCGGCAGTTTAAAGAAATCTACAGGGATGTGCTTAATAGCCTGCCACT
GGACGAAGCAACTATTAATAGGATCGTAGAAGAAGCTAACTATGCTTTTTCACTGAACCGCGAAGTAAT
GCATGATCTGGAGGACCTGATCAAAGCAGCGATAGGCGAGCATACCTTTGATCTGCTTACCAGACAAG
ATCGCCCAGGTTCGACCGAAGCTCGATCGACGGCAGGTCACCCGATTACGCTTATGTTTGTGAATAACT
CGAGTCTGGTAAAGAAACCGCTGCTGCGAAATTTGAACGCCAGCACATGGACTCGTCTACTAGCGCAG
CTTAATTAA

**miRFP670nano3**

*Near-IR FP*

ATGGGGAGCCACCATCACCATCACCATGGCAGATCTGCAGCTGGTACCATGGCAAACCTGGACAAGAT
GCTGAACACCACCGTGACCGAGGTGCGCAAGTTCCTGCAAGCGGACAGAGTGTGCGTGTTCAAGTTC
GAGGAAGATTACTCCGGCACCGTCTCGCACGAAGCCGTGGACGACAGATGGATTAGCATCCTGAAGA
CCCAGGTGCAGGACAGATACTTCATGGAAACCAGAGGCGAGGAATACGTCCACGGCAGATACCAGGC
CATCGCCGACATCTACACAGCCAATCTGGTCGAGTGCTACAGAGACCTGCTGATCGAGTTTCAGGTGC
GGGCCATTCTGGCTGTCCCCATCCTGCAAGGCAAGAAGCTGTGGGGCCTGCTGGTGGCCCACCAACT
GGCCGGCCCTCGGGAGTGGCAGACCTGGGAAATCGACTTCCTGAAACAGCAAGCCGTGGTGATGGG
CATCGCCATCCAGCAGAGCTGA

*DNA (T7 promoter for HO-1)*

CTGAATTCGAAGCTTGGCTGTTTTGGTGGATGAGAGAAGATTTTCAGCCTGATACAGATTAAATCAGA
ACGCAGAAGCGGTCTGATAAAACAGAATTTGCCTGGCGGCAGTAGCGCGGTGGTCCCACCTGACCCC
ATGCCGAACTCAGAAGTGAAACGCCGTAGCGCCGATGGTAGTGTGGGGTCTCCCCATGCGAGAGTAG
GGAACTGCCAGGCATCAGTATACACTCCGCTATCGCTACGTGACTGGGTCATGGCTGCGCCCCGACAC
CCGCCAACACCCGCTGACGCGCCCTGACGGGCTTGTCTGCTCCCGGCATCCGCTTACAGACAAGCTGT
GACCGTCTCCGGGAGCTGCATGTGGAGCTCGGCGCGCCTGCAGGTCGACAAGCTTGCGGCCGCATAA
TGCTTAAGTCGAACAGAAAGTAATCGTATTGTACACGGCCGCATAATCGAAATTAATACGACTCACTATA
GGGGAATTGTGAGCGGATAACAATTCCCCATCTTAGTATATTAGTTAAGTATAAGAAGGAGATATACAT

*HO-1*

ATGACGAATCTTGCGCAGAAACTCAGGTATGGAACCCAACAGAGCCATACCTTGGCAGAGAACACCG
CATACATGAAGTGCTTCCTTAAAGGCATTGTTGAACGTGAACCTTTTAGACAGCTGCTCGCCAATTTGT
ACTACCTTTATTCCGCACTGGAGGCCGCACTCCGGCAGCATCGCGACAATGAGATCATCAGTGCAATCT
ATTTTCCTGAACTGAATCGTACGGACAAGCTCGCAGAGGATCTGACGTATTATTACGGTCCGAATTGGC
AGCAAATCATTCAGCCAACACCGTGCGCCAAAATATATGTAGATCGATTGAAAACGATTGCTGCATCGG
AGCCGGAATTGCTTATCGCTCACTGTTATACCCGCTATCTTGGTGATTTGTCGGGCGGCCAAAGTCTTA
AAAACATTATTCGTTCAGCACTGCAGTTGCCCGAAGGTGAAGGGACGGCTATGTATGAGTTCGACAGC
CTGCCAACACCGGGAGATCGCCGGCAGTTTAAAGAAATCTACAGGGATGTGCTTAATAGCCTGCCACT
GGACGAAGCAACTATTAATAGGATCGTAGAAGAAGCTAACTATGCTTTTTCACTGAACCGCGAAGTAAT
GCATGATCTGGAGGACCTGATCAAAGCAGCGATAGGCGAGCATACCTTTGATCTGCTTACCAGACAAG
ATCGCCCAGGTTCGACCGAAGCTCGATCGACGGCAGGTCACCCGATTACGCTTATGTTTGTGAATAACT
CGAGTCTGGTAAAGAAACCGCTGCTGCGAAATTTGAACGCCAGCACATGGACTCGTCTACTAGCGCAG
CTTAAC

The larger buffer in the DNA between the proteins was required by TwistBiosciences.

**H2B-mRhubarb720**:

ATGCCAGAGCCAGCGAAGTCTGCTCCCGCCCCGAAAAAGGGCTCCAAGAAGGCGGTGACTAAGGC
GCAGAAGAAAGGCGGCAAGAAGCGCAAGCGCAGCCGCAAGGAGAGCTATTCCATCTATGTGTACA
AGGTTCTGAAGCAGGTCCACCCTGACACCGGCATTTCGTCCAAGGCCATGGGCATCATGAATTCGTT
TGTGAACGACATTTTCGAGCGCATCGCAGGTGAGGCTTCCCGCCTGGCGCATTACAACAAGCGCTCG
ACCATCACCTCCAGGGAGATCCAGACGGCCGTGCGCCTGCTGCTGCCTGGGGAGTTGGCCAAGCAC
GCCGTGTCCGAGGGTACTAAGGCCATCACCAAGTACACCAGCAGCTCTAAGGATCCACCGGTCGCC
ACCATGGGCAGCAGCAGCCAGGATCCGATGGCTGAAGGATCCGTCGCCAGGCAGCCTGACCTCTTG
ACCTGCGACGATGAGCCGATCCATATCCCCGGTGCCATCCAACCGCATGGACTGCTGCTCGCCCTCG
CCGCCGACATGACGATCGTTGCCGGCAGCGACAACCTTCCCGAACTCACCGGACTGGCGATCGGCG
CCCTGATCGGCCGCTCTGCGGCCGATGTCTTCGACTCGGAGACGCACAACCGTCTGACGATCGCCTT
GGCCGAGCCCGGGGCGGCCGTCGGAGCACCGATCACTGTCGGCTTCACGATGCGAAAGGACGCAG
GCTTCATCGGCTCCTGGCATCGCCATGATCAGCTCATCTTCCTCGAGCTCGAGCCTCCCCAGCGGGAC
GTCGCCGAGCCGCAGGCGTTCTTCCGCCACACCAACAGCGCCATCCGCCGCCTGCAGGCCGCCGAA
ACCTTGGAGAGCGCCTGCGCCGCCGCGGCGCAAGAGGTGCGGAAGATTACCGGCTTCGATCGGGT
GATGATCTATCGCTTCGCCTCCGACTTCAGCGGCGAAGTGATCGCAGAGGATCGGTGCGCCGAGGT
CGAGTCAAAACTAGGCCAGCACTATCCTGCCTCAGATATTCCGGCGCAGGCCCGTCGGCTCTATACC
ATCAACCCGGTACGGATCATTCCCGATATCAATTATCGGCCGGTGCCGGTCACCCCAGACCTCAATC
CGGTCACCGGGCGGCCGATTGATCTTAGCTTCGCCATCCTGCGCAGCGTCTCGCCCGTCCATCTGGA
ATTCATGCGCAACATAGGCATGCACGGCACGATGTCGATCTCGATTTTGCGCGGCGAGCGACTGTG
GGGATTGATCGTTTGCCATCACCGAACGCCGTACTACGTCGATCTCGATGGCCGCCAAGCCTGCGAG
CTAGTCGCCCAGGTTCTGGCCCGGGCGATCGGCGTGATGGAAGAGTAA

**Section S2 Protein production and purification**

Plasmids containing the near-IR FPs (pET-DUET-1-mRhubarb720-HO [Addgene: #141201]; custom pET28a+ vectors for emiRFP670, miRFP670nano3 with HO1 (TwistBiosciences) (see Section S1 for gene sequences for each construct) were used. Near-IR constructs were transformed into chemically competent *E.coli* BL21(DE3) (New England Biolabs). For mCherry, a pBAD expression system was used [Addgene: #54630]; transformed into chemically competent *E.coli* TOP10 (New England Biolabs). All constructs were plated on LB agar ampicillin (50 μg/ml) and incubated at 37°C overnight. Selected colonies were then picked and incubated in 5 mL 2xYT ampicillin (50 μg/ml) cultures overnight at 37°C. The 5 ml overnight cultures were then used to inoculate 0.5 L 2xYT ampicillin (50 μg/ml) cultures. For the IRFPs, cultures were incubated at 37°C until an $OD_{600}$ of 0.4-0.8 was reached, upon which 0.5 mM IPTG was added to induce IRFP expression. The cultures were then incubated for a further 24 h at 22°C with shaking. For mCherry, cultures were incubated at 37°C until an $OD_{600}$ of 0.4-0.8 was reached, upon which 1% (w/v) L-Arabinose was added to induce mCherry expression. These cultures were then incubated for a further 18 h at 25°C with shaking.

Harvested cells were resuspended in 50 mM Tris buffer (pH 8.0) and lysed using a French press; cell lysate was clarified by centrifugation and supernatant was collected. The FPs were purified using a His-trap HP nickel affinity column (GE Healthcare) in conjunction with ATKA Purifier. The His-trap column was equilibrated with 50 mM Tris (pH 8.0), 300 mM NaCl and 10 mM Imidazole and eluted with 50 mM Tris (pH 8.0), 300 mM NaCl with 500 mM Imidazole. Pooled FPs samples were then subjected to size exclusion chromatography using a HiLoad 26/600 Superdex 200 column, where the column was equilibrated with 50mM Tris (pH 8.0). Elution was monitored as specific wavelengths, depending on the $\lambda_{max}$ of the FP. Protein purity was validated using SDS-PAGE analysis.

**Section S3 Protein electronic excitation characterisation**

Each purified sample was subject to a DC Protein Assay (BioRAD) to determine protein concentrations using the manufactures guidelines. BSA (0.2-1.5 mg/ml) was used as the concentration standard. UV-Vis spectroscopy was performed using a Cary 60 spectrophotometer (Agilent Technologies). Characterisations used 10 μM of each sample for absorbance measurements across 200–800 nm. Extinction coefficients were calculated using the Beer-Lambert law.

**Table S1.** Spectral properties of IRFPs used in this study.

| IFRP | Determined here | | Previously reported [a] | |
|---|---|---|---|---|
| | $\lambda_{max}$ (nm) | $\varepsilon$ ($mM^{-1}cm^{-1}$) | $\lambda_{max}$ (nm) | $\varepsilon$ ($mM^{-1}cm^{-1}$) |
| mRhubarb720 | 703 | 72.5 | 701[b] | 95.0[b] |
| emIRFP670 | 644 | 23.2 | 642[c] | 87.4[c] |
| miRFP670nano3 | 645 | 31.9 | 645[d] | 129.0[d] |

a, values reported in FPbase (https://www.fpbase.org/) [1]
b, Reported by Rogers et al. [2]
c, Reported by Matlashov et al. [3]
d, Reported by Oliinyk et al. [4]

**Section S4 Crystallography**

Crystallisation screens were set up using hanging-drop vapor diffusion method for the purified mRhubarbWT at 10 mg/mL, with 0.1 M MMT buffer (pH 5.0) and 25% (w/v) polyethylene glycol 1500. Micro-crystals of mRhubarb grew which were suitable for the microfocus macromolecular crystallography beamline VMXm [5] at Diamond Light Source. Crystals were prepared on cryo-electron microscopy grids, using a Leice EM GP2 plunge freezer, with vitrification in liquid ethane, before mounting on the VMXm beamline. 30° wedges of data were collected from multiple crystals, with an oscillation of 0.1° and exposure time of 0.2 s per image, with 100% transmission of beam. The data were collected at an energy of 21.326 keV (wavelength of 0.5814 Å). Each individual dataset was processed with DIALS [6], with the resulting processed data being input into xia2.multiplex [7] to merge and produce a complete dataset.. Initial structure determination was carried out by Alphafold [8] with manual molecular replacement for the Biliverdin XI-alpha cofactor using CCP4 [9] and COOT.[10] The refined structure of mRhubarbWT was visualised using PyMOL. Data collection and refinement statistics can be found in Table S2.

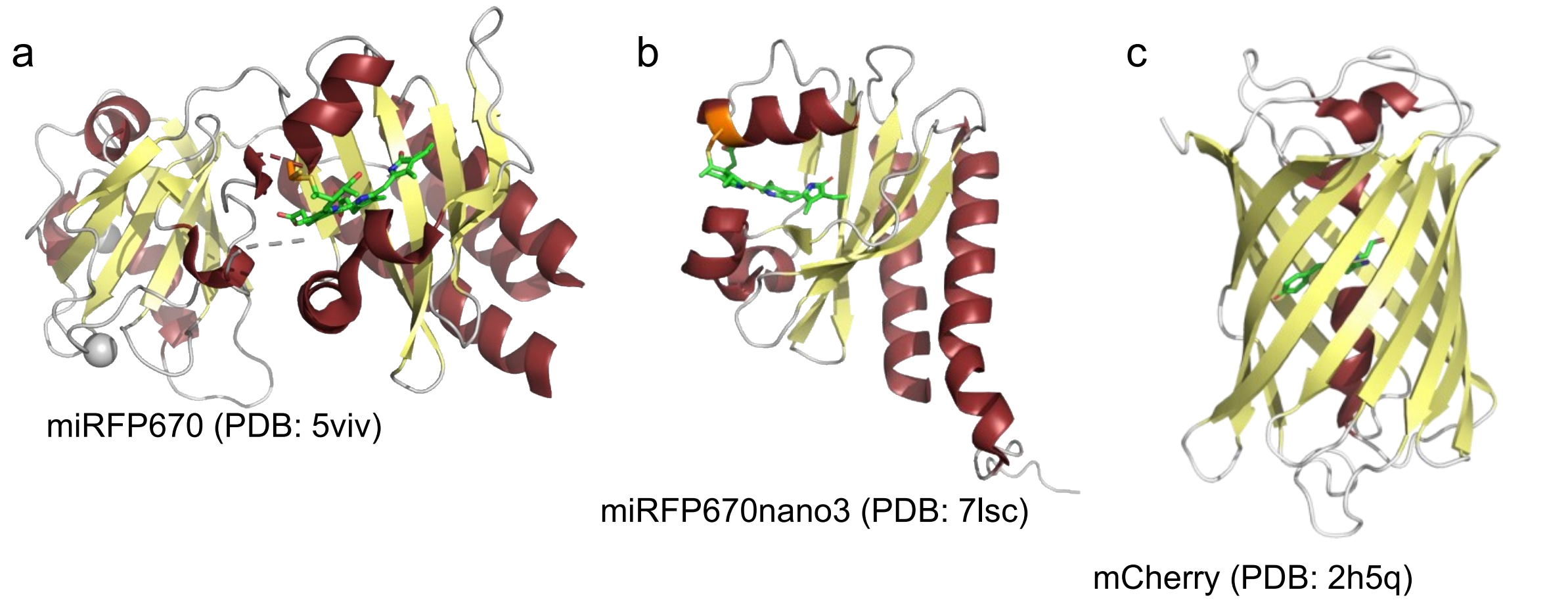

**Figure S1.** Structures of various infrared and red fluorescent proteins. (a) miRFP670 (PDB 5viv)[11], an engineered truncated IFRP miRFP670nano3 (PDB 7lsc)[4] and the β-barrel-type fluorescent protein with the autocatalytically forming chromophore, mCherry (PDB 2h5q)[12].

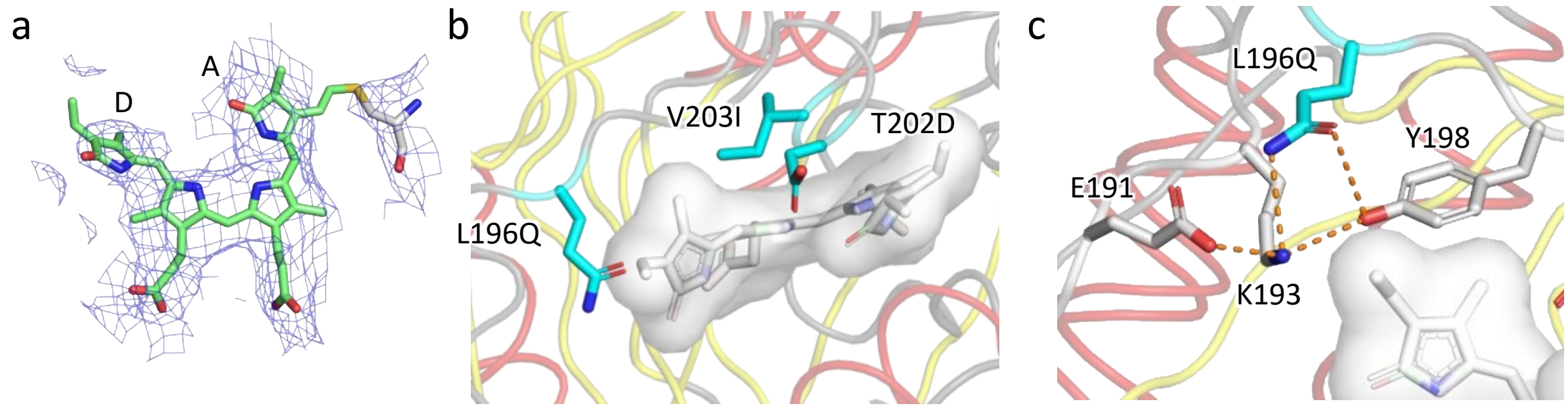

**Figure S2.** mRhubarb720 structural details. (a) the $2F_0$-$F_c$ map of biliverdin set to 1.5 σ. The thioether link between ring A and C15 is shown for reference. (b) The positioning of key mutations generating mRhubarb720. (c) Potential polar interaction network between L196Q and local residues. We note the lack of electron density around the supposed thioether bond.

This could indicate that the bond is flexible and prone to movement, and/or the presence of unbound BV in the crystals measured.

**Table S2.** Data collection and refinement statistics for mRhubarb
* Three crystals were merged to produce this data set.
[1] Coordinate Estimated Standard Uncertainty in (Å), calculated based on maximum likelihood statistics.

| PDB Entry | 9F03 |
|---|---|
| **Data Collection*** | |
| Diamond Beamline | VMXm |
| Date | 21 Jan 2022 |
| Wavelength (Å) | 0.58136 |
| **Crystal Data (figures in brackets refer to outer resolution shell)** | |
| a,b,c (Å) | 58.85, 58.85, 185.88 |
| a,b,g (°) | 90.0, 90.0, 90.0 |
| Space group | $P\ 4_1\ 2_1\ 2$ |
| Resolution (Å) | 3.30 – 27.41 |
| Outer shell | 3.30 – 3.57 |
| R-merge (%) | 33.4 (164.5) |
| R-pim (%) | 20.2 (106.7) |
| R-meas (%) | 37.2 (179.3) |
| CC1/2 | 0.971 (0.315) |
| I / σ(I) | 4.8 (0.7) |
| Completeness (%) | 95.1 (90.7) |
| Multiplicity | 4.2 (2.9) |
| Total Measurements | 21,465 (2,786) |
| Unique Reflections | 5,072 (953) |
| Wilson B-factor(Å$^2$) | 77.6 |
| **Refinement Statistics** | |
| Refined atoms | 2,382 |
| Protein atoms | 2,339 |
| Non-protein atoms | 43 |
| Water molecules | 0 |
| R-work reflections | 4,779 |
| R-free reflections | 261 |
| R-work/R-free (%) | 22.5 / 30.9 |
| **rms deviations (ML target in brackets)** | |
| Bond lengths (Å) | 0.008 (0.013) |
| Bond Angles (°) | 1.448 (1.648) |
| [1]Coordinate error | 0.892 |
| Mean B value (Å$^2$) | 86.0 |
| **Ramachandran Statistics (PDB Validation)** | |
| Favoured/allowed/Outliers | 230 / 46 / 28 |
| % | 75.7 / 15.1 / 9.2 |

**Section S5 SRS, DIC, and fluorescence microscopy**

Measurements were acquired using a home-built coherent Raman microscope setup described in our previous work[13] with simultaneous SRS, forward and epi confocal-fluorescence detection. A sketch of the set-up is shown in Fig. S3. SRS is measured in the form of stimulated Raman loss (SRL) at a pump wavelength of 820 nm. For SRS, the Stokes beam is amplitude modulated at 2.5 MHz using an acoustic-optic modulator (AOM). Pump and Stokes are recombined using a dichroic beam splitter (DBS1, Melles Griot LWP-45-RP808-TP1064-PW-1025-C) and raster-scanned on the sample using a galvo scanner and a 4f optical relay from the scanner to the microscope objective (MO), mounted onto a commercial inverted microscope stand (Nikon Ti-U). The signal is collected by a 1.34 NA oil-immersion condenser (Nikon MEL41410) in transmission geometry. SRL is detected as an intensity modulation $\Delta I/I$ by a photodiode (PD, Hamamatsu S6976) using a short-pass filter F2 (Semrock FF01-945/SP and Thorlabs DMLP900R in reflection) to separate the pump from the Stokes beam, a home-built resonant preamplifier at 2.5 MHz, and a lock-in amplifier (Zurich Instruments HF2LI). When using IFD centre at 1640cm$^{-1}$, the short-pass filter F2 was replaced with a stack of two Thorlabs FESH0850, to allow for Stokes wavelengths closer to the pump.

A dichroic beamsplitter (DBS2, Semrock FF776-Di01) in the detection was used to transmit/reflect signal above/below 776 nm, to separate SRL from fluorescence. The fluorescence signal was spectrally selected by using a bandpass filter F1 (Semrock FF01-675/67) transmitting 642-707 nm and detected by a photomultiplier (PMT- Hamamatsu H10770A-40). Note that this forward PMT detection is also sensitive to coherent anti-Stokes Raman scattering (CARS) in the wavenumber range above 1750 cm$^{-1}$, as observed for measurements carried out on the Tris buffer (Fig. S6). Epi-fluorescence in the range 560-750 nm is reflected by the dichroic beamsplitter DBS3 (Eksma) and filtered by F3 (Semrock FF02-650/100) transmitting 600-700 nm, focussed through an adjustable confocal pinhole PH, and detected by a photomultiplier E-PMT (Hamamatsu H10770-40). The PH was adjusted to a diameter of twice the Airy disk of the objective (equivalent to $2\times1.22\times650$ nm/NA at the sample).

Spectral focussing is implemented using glass blocks as described in our previous work.[13] Here, several 11 cm long H-ZF88 glass blocks were added to the setup to increase the maximum group delay dispersion and achieve a spectral resolution down to 6.7 cm$^{-1}$. The longer pulses used to achieve higher spectral resolution also reduce the signal, as is seen in Fig. S5, where SRS on a polystyrene (PS) calibration sample and mRhubarb720 is shown for 6.7, 11, and 21 cm$^{-1}$ resolution. Improving the spectral resolution from 21cm$^{-1}$ to 11 cm$^{-1}$ results in better resolved mRhubarb720 resonances. Conversely, further decreasing to 6.7 cm$^{-1}$ mostly reduces the signal, indicating that the resonance linewidth is larger than 6.7 cm$^{-1}$. Therefore, we used 11 cm$^{-1}$ resolution for the measurements shown in this work. This resolution was obtained by adding 160 mm of H-ZF52A and 220 mm H-ZF88 in the pump beam, stretching the pulses from 106 fs to 3.1 ps intensity full-width at half-maximum (FWHM), and 160 mm of H-ZF52A and 330 mm H-ZF88 in the Stokes beam stretching the pulse from 129 fs to 2.7 ps.

The centre IFD was 1740 cm$^{-1}$ using a Stokes wavelength of 956 nm. After replacing the filter F2, the centre IFD was 1640 cm$^{-1}$ using a Stokes wavelength of 948 nm. Choosing the centre at 1640cm$^{-1}$ positions the 1620cm$^{-1}$ vibrational resonance below this centre IFD value. This leads to a higher CRS signal in spectral focussing[14] due to the time ordering with the Stokes arriving slightly earlier at resonance. By increasing the centre IFD, the vibrational resonant signal was eventually lost, when the required IFD, obtained by changing the relative delay time between pump and Stokes pulses, was outside of pump-Stokes time overlap. Pump-Stokes two-photon absorption (TPA) manifests as a pump-loss in the SRL detection, and is clearly visible as a pump-Stokes temporal cross-correlation pulse intensity profile, since changing wavenumber equates to tuning the pump-Stokes relative time delay. This is shown in Fig. S10, plotting the SRL spectrum for mRhubarb720 measured at different centre IFDs. After changing the filter F2, lower centre IFD could be reached, and Fig. S8 shows the corresponding SRL spectra of mRhubarb720. The signal increased by about an order of magnitude using 1640 $cm^{-1}$ with a Stokes wavelength of 948 nm, due to the more efficient resonant driving at pump-Stokes overlap.

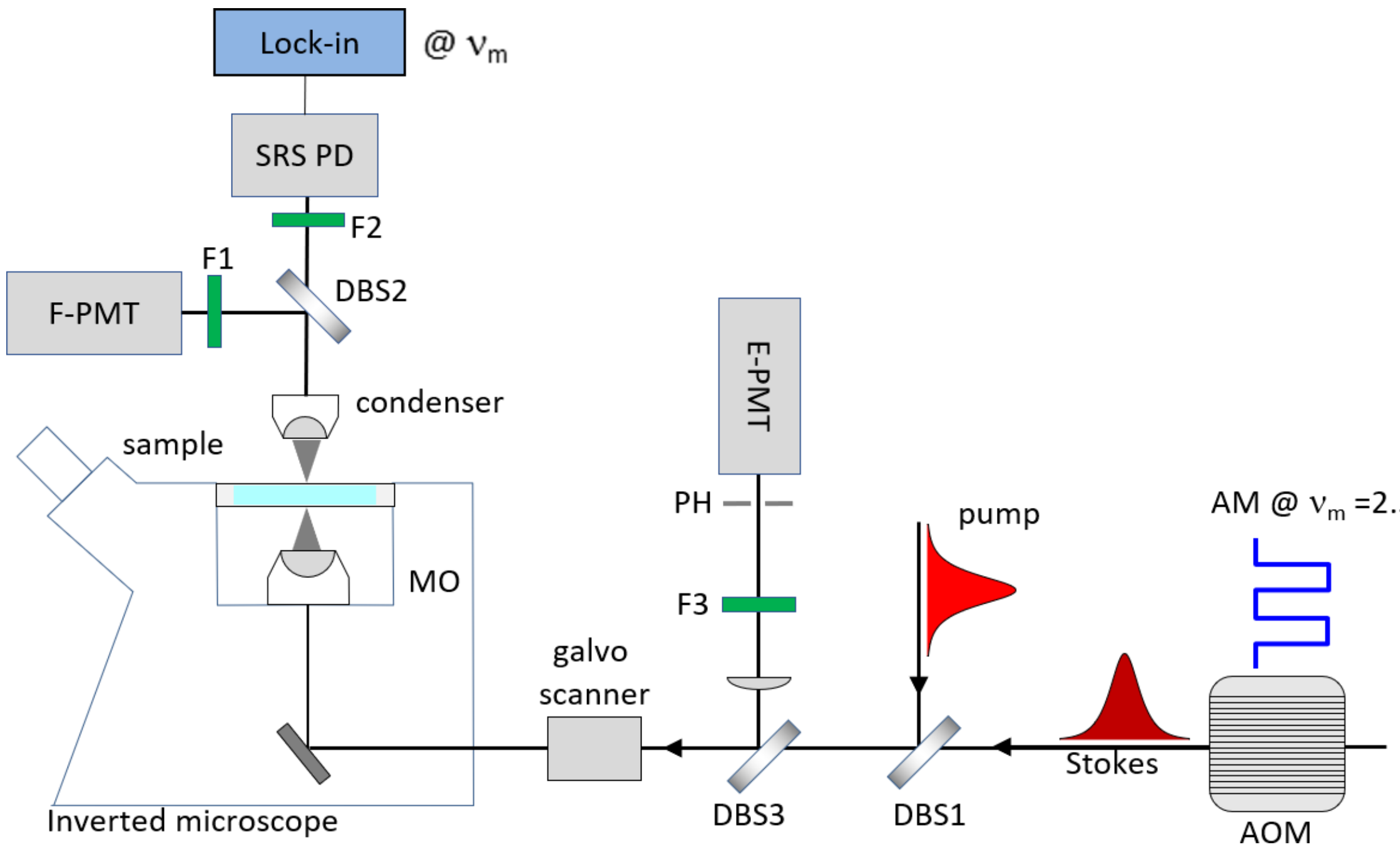

**Figure S3.** SRS microscopy set-up. The Stokes beam is amplitude modulated (AM) at 2.5 MHz using an acoustic-optic modulator (AOM). Pump and Stokes are recombined using the dichroic beam splitter DBS1 and raster scanned onto the sample using a galvo scanner. They are focused using a microscope objective (MO) mounted onto a commercial inverted microscope stand and the signal is collected by a 1.34 NA oil-immersion condenser in transmission geometry. SRL is detected as an intensity modulation by a photodiode (PD) using a short-pass filter F2 to separate the pump from the Stokes beam, a home-built resonant circuit at 2.5 MHz, and a lock-in amplifier. A dichroic beamsplitter DBS2 in detection is used to transmit/reflect signal above/below 776 nm, to separate SRL from fluorescence. The fluorescence signal is spectrally selected by using a bandpass filter F1 and detected by a photomultiplier F-PMT. Epifluorescence in the range 560-750 nm is reflected by the dichroic beamsplitter DBS3 filtered by F3, focussed through the adjustable confocal pinhole PH, and detected by a photomultiplier E-PMT.

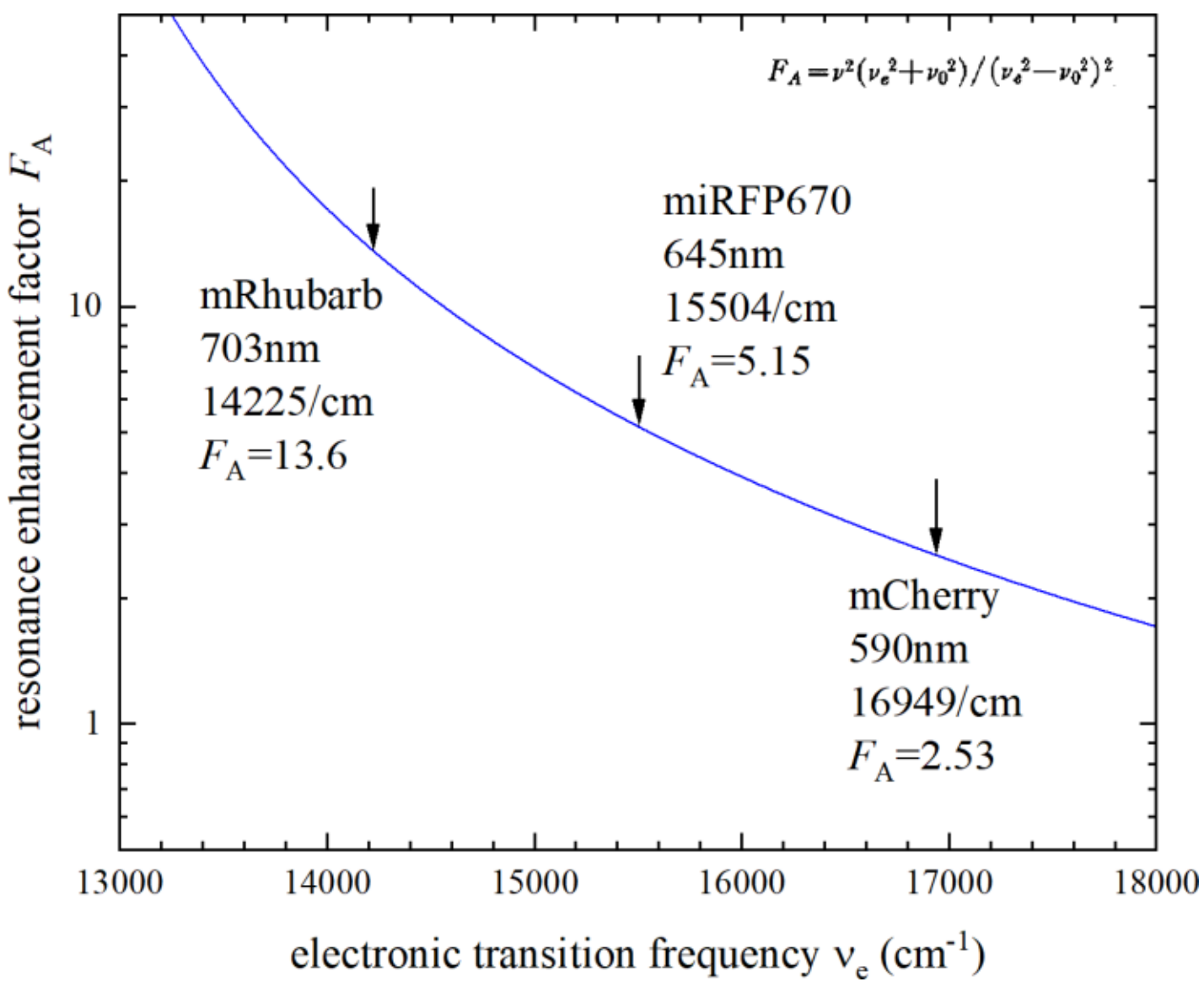

**Figure S4.** Resonance enhancement factor $F_A = (\nu_p - \nu_v)^2 (\nu_e^2 + \nu_p^2) (\nu_e^2 - \nu_p^2)^{-2}$ (see Eqn. 5 in [15]) for a pump wavelength of 820 nm and vibrational frequency of $\nu_v$=1650 cm$^{-1}$**.** The electronic transition frequencies $\nu_e$ for the fluorophores measured are shown.

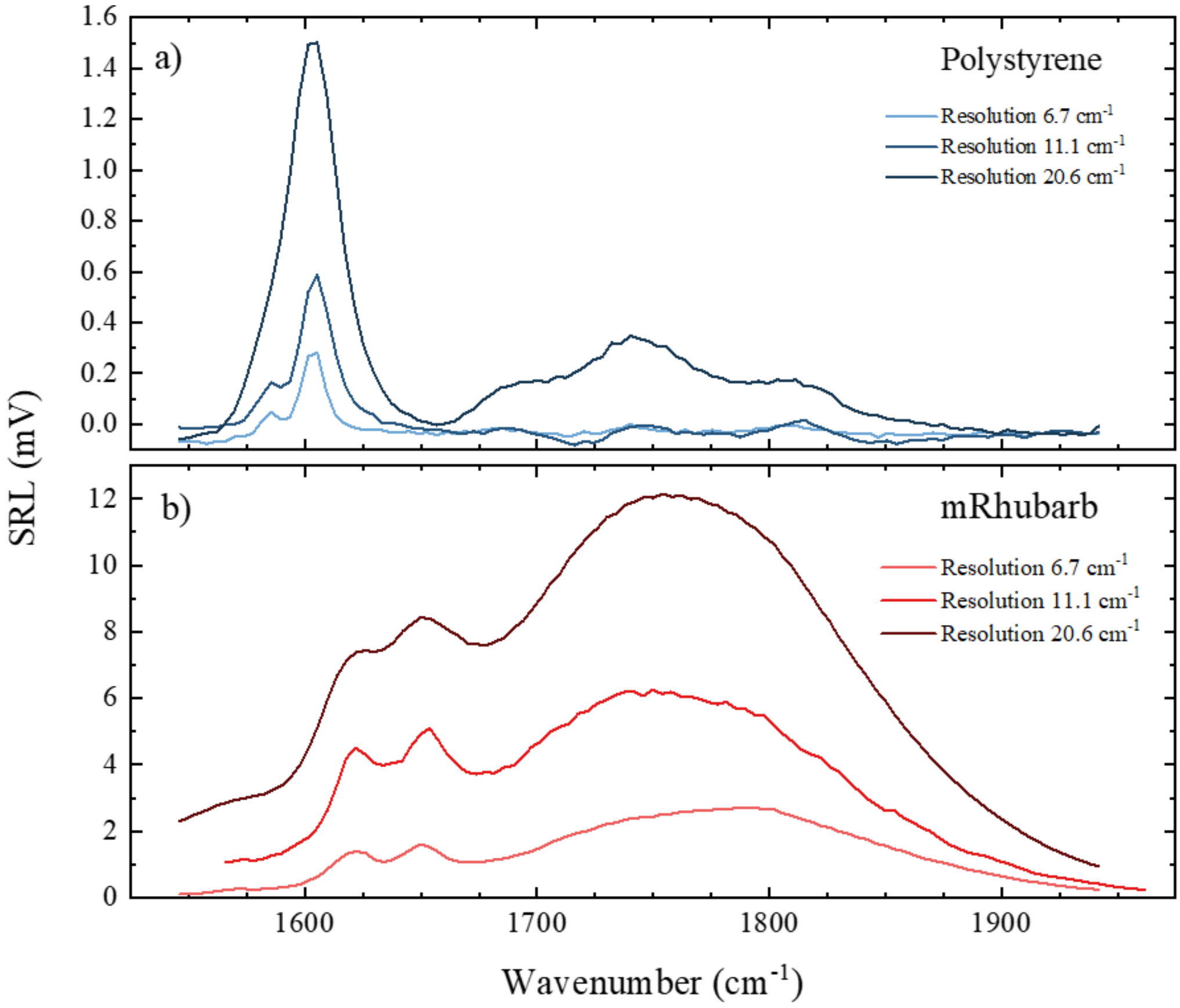

**Figure S5.** Comparison of the SRS spectra of a) polystyrene and b) mRhubarb720 in solution (1 mM), measured at different spectral resolutions, with 6 mW pump and 8 mW Stokes power at the sample, using the 20× 0.75 NA objective, and centre IFD 1740 cm$^{-1}$.

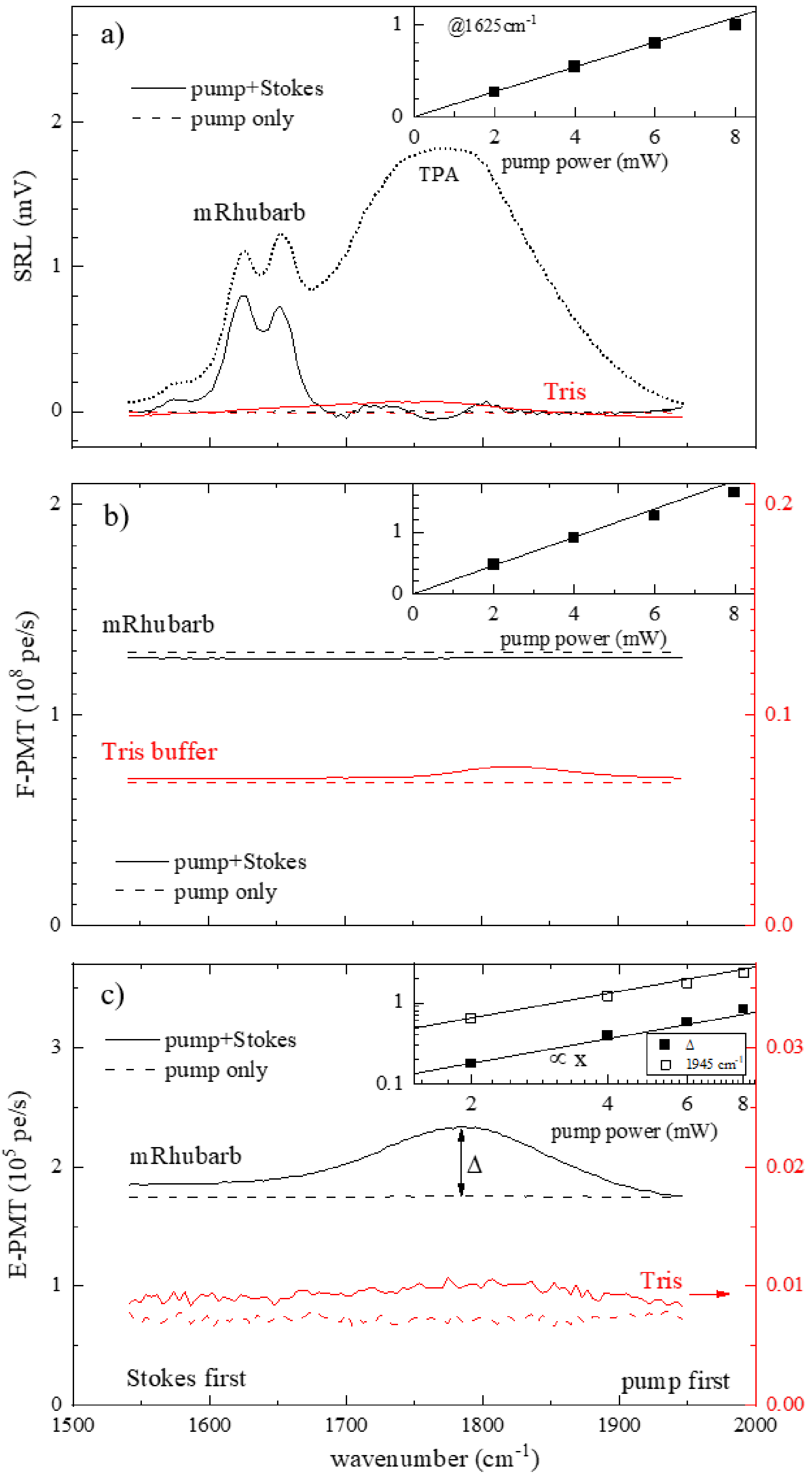

**Figure S6**. Comparison between SRL (a), forward PMT (b) and confocal epi-PMT (c) signals for mRhubarb720 in solution (1 mM) at 6 mW pump and 8 mW Stokes power at the sample, 20× 0.75 NA objective. Insets show dependence on pump-power of (a) vibrationally resonant

response, (b) pump + Stokes, (c) Stokes. Experimental conditions are the same as in Fig. 2. The results without the Stokes beam (dashed), and for the Tris buffer (red) are also shown.

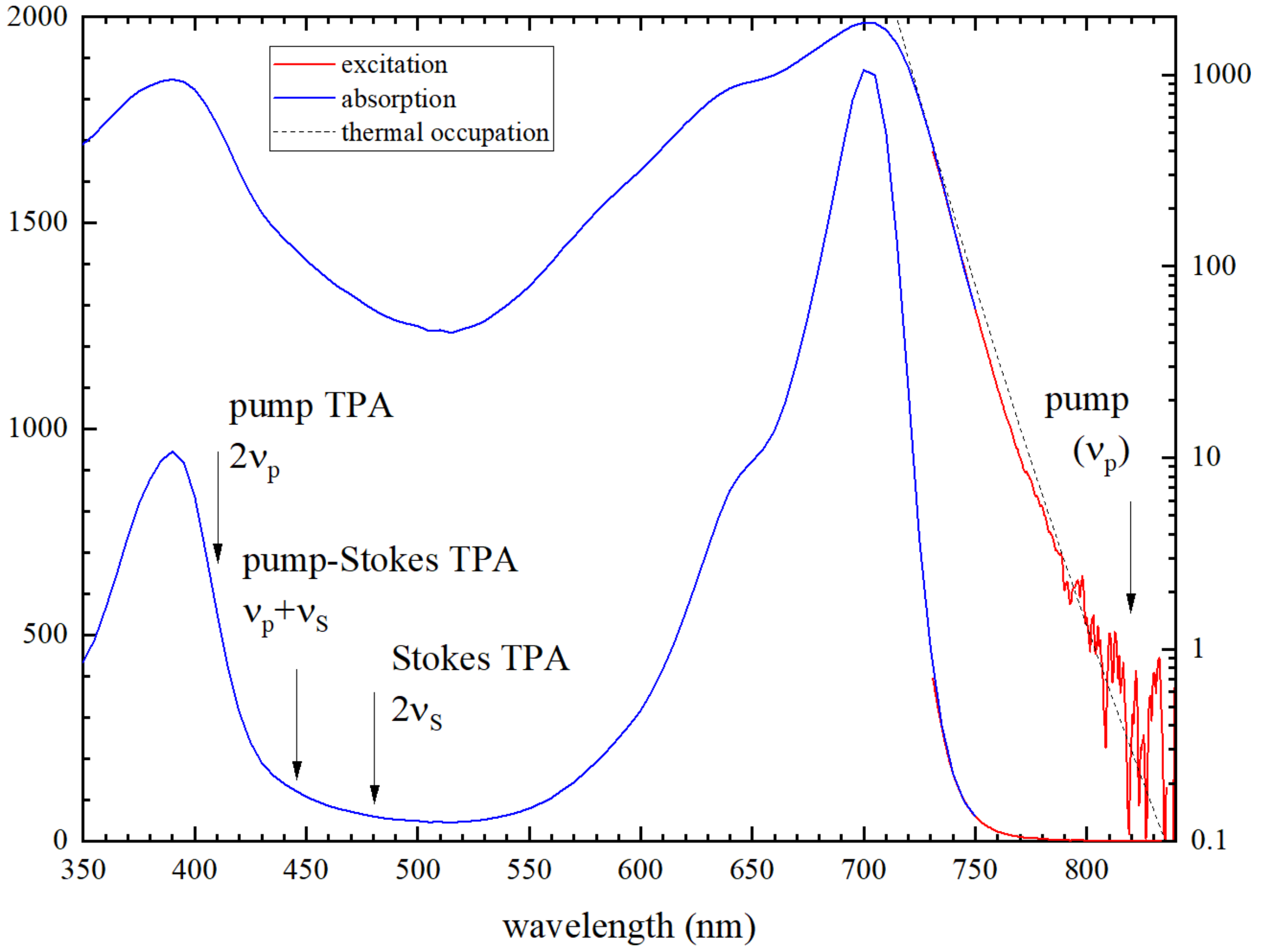

**Figure S7.** Absorption (blue) and fluorescence excitation spectra (red) of mRhubarb, on a linear and logarithmic scale, using arbitrary units. The absorption and excitation spectra have been scaled relatively, to match around 750 nm, with the excitation spectra allowing to show the absorption profile tail with 4 orders of magnitude dynamic range. A thermal occupation curve proportional to $\exp(h\nu/(k_B T))$ with Plack's constant $h$, Boltzmann's constant $k_B$ and the absolute temperature $T = 293\,\mathrm{K}$ is plotted as dashed line, showing good agreement with the absorption tail. The wavelength position of TPA from pump, Stokes and combined pump-Stokes photons are indicated.

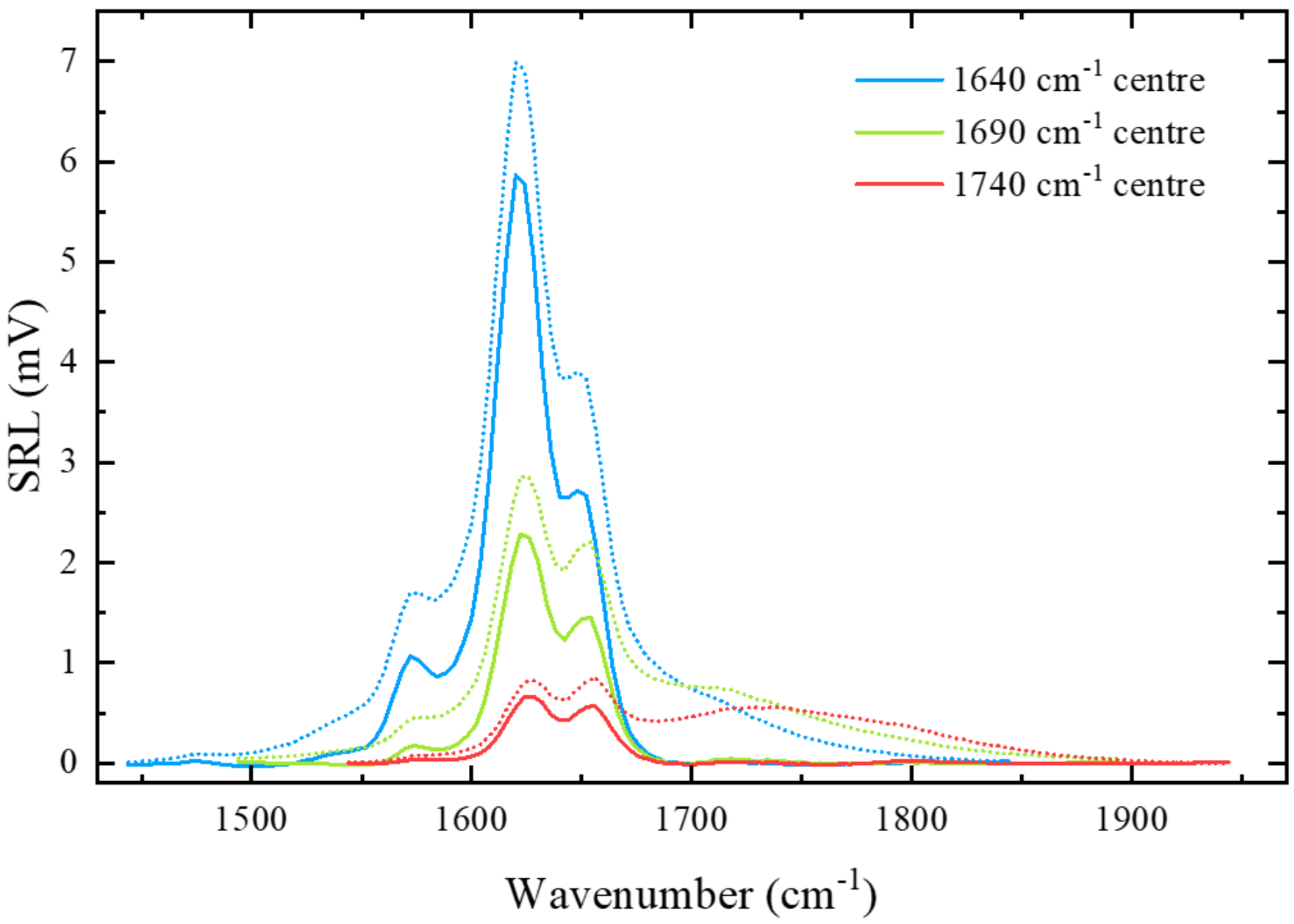

**Figure S8.** SRL signal of mRhubarb720 in solution (1 mM, with 6 mW pump and 8 mW Stokes power at the sample, 20× 0.75 NA objective, 11 $cm^{-1}$ resolution) measured at different centre IFDs, as indicated, by using Stokes centre wavelengths of 956, 952 and 948 nm. Dotted lines as measured, solid lines TPA background subtracted.

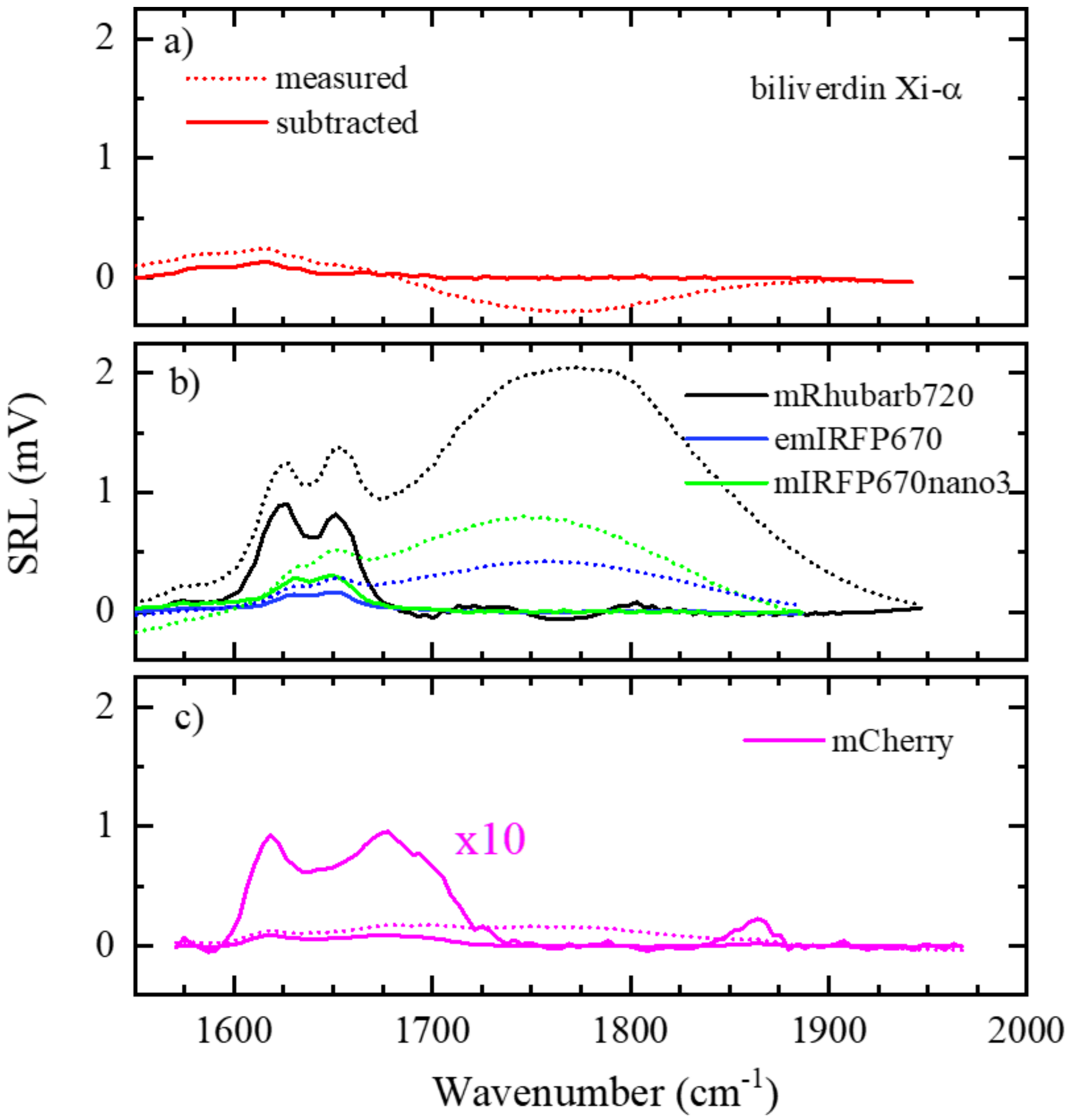

**Figure S9.** IRFPs SRL spectra before (dotted) and after (solid) subtraction of the pump-Stokes TPA effect. Same experimental conditions as in Fig. 2.

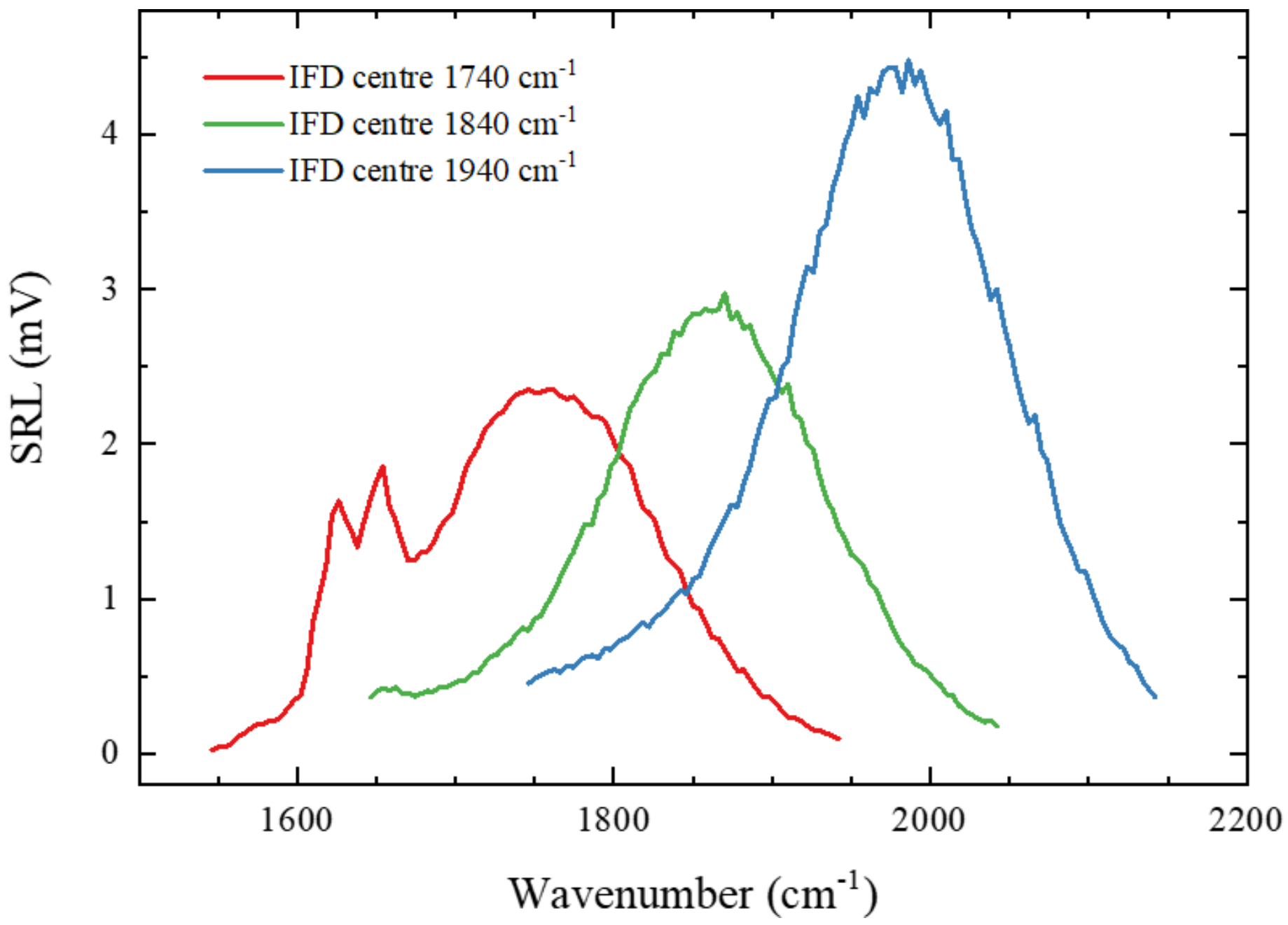

**Figure S10.** SRL signal of mRhubarb720 in solution (1 mM, with 6 mW pump and 8 mW Stokes power at the sample, 20× 0.75 NA objective, 11 $cm^{-1}$ resolution) measured at different centre IFDs as indicated using Stokes centre wavelengths of 956, 966 and 975 nm. TPA results in an apparent SRL signal that follows the pump-Stokes temporal overlap (different wavenumbers correspond to different pump-Stokes relative delay times in spectral focussing).

To measure SRL spectra of the IRFPs in solution, we added 1 mM in 50 mM Tris buffer pH 8 into Grace BioLabs CultureWell™ gaskets having wells of 3 mm diameter and 1 mm depth, mounted between a slide and coverslip. A 20× 0.75 NA microscope objective (Nikon MRD00205) and 1.5 × tube lens was used, with the focus close to the centre of the well. SRL was measured for pump-Stokes delay scans over an IFD range of 1540 – 1940 $cm^{-1}$ using 101 samples and 4 $cm^{-1}$ step size. At each IFD, a spatial line scan over 10 µm with 1500 samples and 0.1 ms pixel dwell time was taken. The scans were then averaged over twenty repetitions and over the spatial dimension (corresponding to 3 s aggregated acquisition time per spectral point) to produce a plot of SRL signal against wavenumber. Pump (Stokes) power at the sample was 6 mW (8 mW) unless otherwise noted, with the Stokes power referring to 50% duty cycle modulated Stokes. An overview of the measured SRL for IRFPs in solution is shown in Fig. 2 and Fig. S9.

Simultaneous acquisition of SRL, forward and epi-detected fluorescence is shown for mRhubarb720 in Fig. S6. The forward-detected fluorescence (F-PMT, Fig. S6b) is strong, around $140 \times 10^6$ photoelectron/s (140 Mph/s), dominated by the emission due to the absorption of the pump-only beam, and scales linearly with the pump-power. It is therefore due to residual one-photon absorption owing to the pre-resonant condition as discussed in detail in section S7. While the pump beam has a frequency well below the absorption peak, at room temperature there is enough thermal population of vibrational states to enable one-photon absorption of the pump into the first electronic excited state, and subsequent fluorescence. The fluorescence detected by the F-PMT is dominated by the contribution from bulk volume in solution, since it is about 3 orders of magnitude higher than the confocal fluorescence detected by the E-PMT shown in Fig. S6c. The latter shows a pump-Stokes two-photon absorption effect which follows the pulse temporal cross-correlation profile while changing the delay time between pump and Stokes pulses in spectral focussing. This TPA-induced fluorescence (see $\Delta$ in Fig. S6c) is superimposed onto a pump-only contribution which scales linearly with the pump power, and thus is again dominated by a one-photon absorption as mentioned above. The lack of a significant pump-only TPA-induced fluorescence is due to its lower power and possibly a lower TPA cross-section for the pump as compared to the pump-Stokes case. Note that we do not see stimulated Raman excited fluorescence (SREF), since this should exhibit vibrational resonances versus wavenumbers[16] which are not observed in the detected fluorescence.

Control measurements without mRhubarb720 (Tris buffer only) show a weak SRL (see Fig. S6a), from spectrally broad vibrational resonances of water. The forward fluorescence contribution (see Fig. S6b) is also very small, ~20 times lower than with mRhubarb720. In this detection channel, there is a small contribution (around 1 Mpe/s) of non-resonant CARS, which is peaked around 1820 $cm^{-1}$, shifted compared to the centre IFD of 1740 $cm^{-1}$ due to the filter cut-off supressing CARS detection for IFDs below 1800 $cm^{-1}$. The epi-detected confocal fluorescence (see Fig. S6c) is very weak (around 1 kpe/s), as there is no CARS in epi-direction for a homogeneous liquid.

As mentioned, the TPA of combined pump - Stokes results in a broad positive background SRL signal. To remove this background, the TPA profile is fitted using least squares to the data in the Curve Fitting Toolbox of MATLAB R2021a. We use a hyperbolic secant pulse shape typical for Kerr-lens mode locked laser pulses, superimposed to a sloped background, as given by

$$y = a \operatorname{sech}^2\left(\frac{x-b}{c}\right) + d + ex \,,$$

where *a*, *b*, and *c* are the amplitude, centre, and width of the pulse, and *d* and *e* are offset and slope of the background. Spectral regions showing strong Raman peaks were excluded from the fit, to isolate the TPA. For biliverdin, a weaker background with a bipolar shape was observed, likely due to a dominating Kerr lensing effect, and fitted using the function

$$y = a \tanh\left(\frac{x-b}{c}\right) + d + ex + f \tanh\left(\frac{x-g}{h}\right).$$

We note that we detect both the in-phase and in-quadrature components of the modulated SRL signal. In-quadrature components are signatures of long-lived responses on the timescale of the modulation period (400 ns), which can be of photothermal origin. We find a negligible signal, about two orders of magnitude below the in-phase signal, close to the noise level, showing that the signals observed here are not significantly affected by photothermal effects.

SRL and quantitative differential interference contrast (qDIC) images in cells were taken using a 60× 1.27 NA water immersion objective (Nikon MRD70650) and the 1× tube lens of the Ti-U inverted microscope stand. SRL was sequentially acquired as individual images at 1620 $cm^{-1}$ and 1660 $cm^{-1}$, and this sequence was repeated 20 times. Pixel dwell times were 0.01 ms. Excitation powers at the sample were typically around 7 mW for pump and 12 mW for Stokes.

The microscope was also equipped with wide-field epi-fluorescence excitation and detection. As excitation, we used a red LED centred at 634 nm having 21 nm width, delivered via a 3 mm diameter lightguide (Bluebox niji 7). In detection, we used a filter cube with an excitation filter (Edmund Optics 67-036) transmitting 603 to 678 nm, a 685 nm dichroic (Edmund Optics 67-085) and an emitter filter (Edmund Optics 87-758) transmitting 698 to 766 nm. The camera was an sCMOS (PCO.edge 5.5 RS) with 2560 x 2160 pixels. Data were taken using the maximum 2 s integration time, slow scan readout having a read noise of 1.6 e, and a gain of 0.5754 e/count. The camera was mounted at the eyepiece port of the Ti-U microscope.

qDIC images were acquired with the same camera, averaged over 128 acquisitions in fast scan readout, using a 100 W halogen lamp with a Thorlabs FB550-40 filter for transillumination. A de-Sénarmont compensator (T-P2 DIC Polariser HT MEN51941, Nikon), was used to control the phase offset, and a N2 DIC module (MEH52500, Nikon) was added in the condenser (HNAO MEL41410, Nikon) of 1.34 numerical aperture (NA), combined with a N2 DIC slider (MBH76190, Nikon) and a Ti-A-E DIC analyser block (MEN 51980, Nikon). The quantitative phase difference images were acquired and analysed as detailed previously[17,18] at a polariser angle of ±30°.

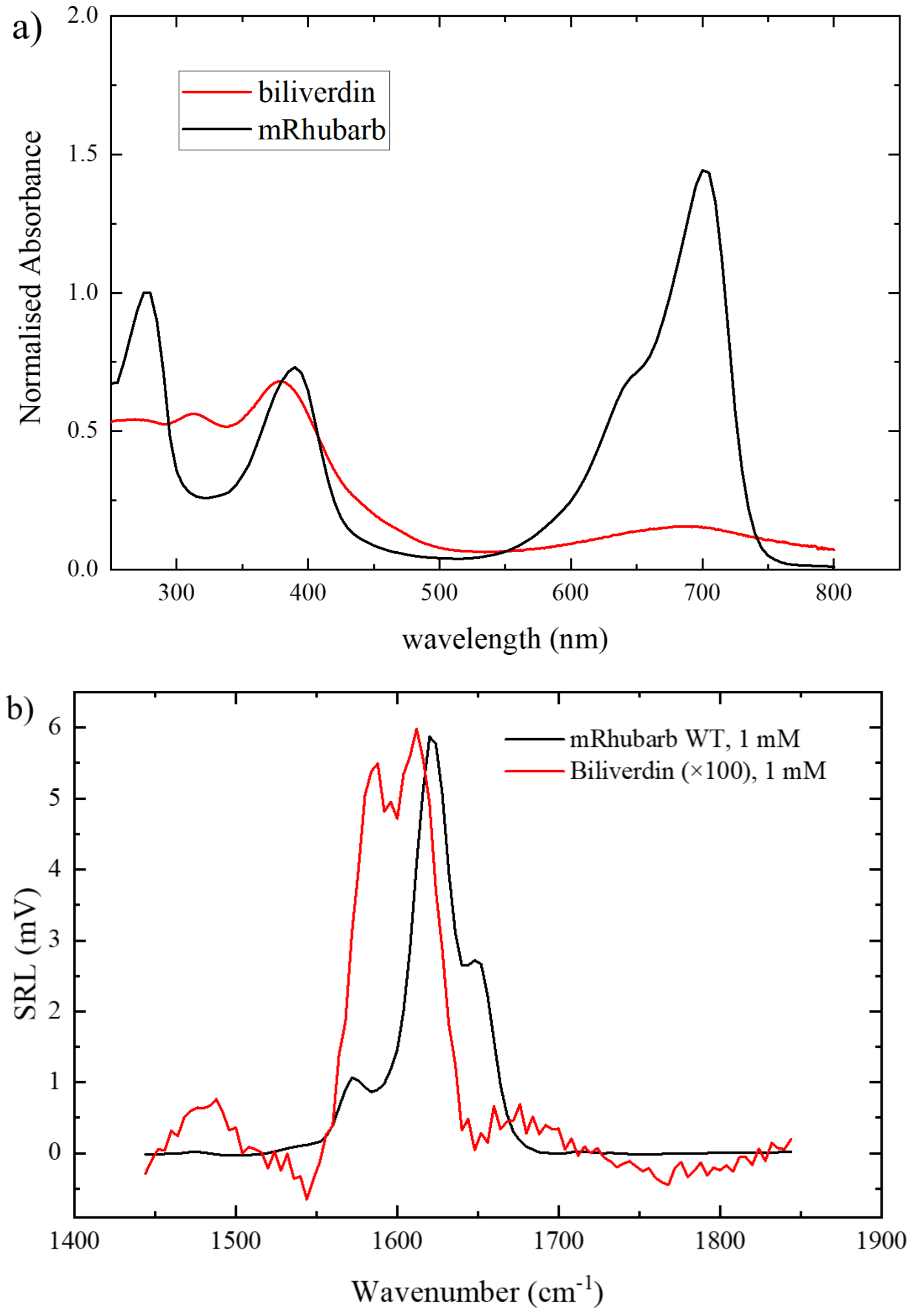

**Figure S11.** Comparison between free BV in 50 mM Tris and 50 mM NaOH (red) and BV bound to the protein component in mRhubarb720 in 50 mM Tris (black). a) Absorption spectrum b) SRL of 1mM free BV scaled by a factor of 100, and 1mM mRhubarb720. Microscope objective 20× 0.75 NA, pump (Stokes) power at the sample 6 mW (15 mW), 0.1 ms pixel dwell time, averaged to 3 s per spectral point, 1640cm$^{-1}$ centre IFD.

**Section S6 Mammalian cell preparation**

For mammalian cell imaging, HeLa cells (ATCC CCL-2) were grown on 25mm #1.5 borosilicate glass coverslips placed in 6 well tissue culture plates. Transfection was achieved using Fugene6 (Promega) according to manufacturer's instructions and allowed to express for 24hours. Coverslips were then washed in PBS before being fixed with 4% paraformaldehyde in PBS for 20 minutes. Following fixation, coverslips were washed in PBS and mounted on a cut down microscope slide using a 0.12 mm imaging gasket (Grace Biolabs) with PBS buffer in the resulting chamber.

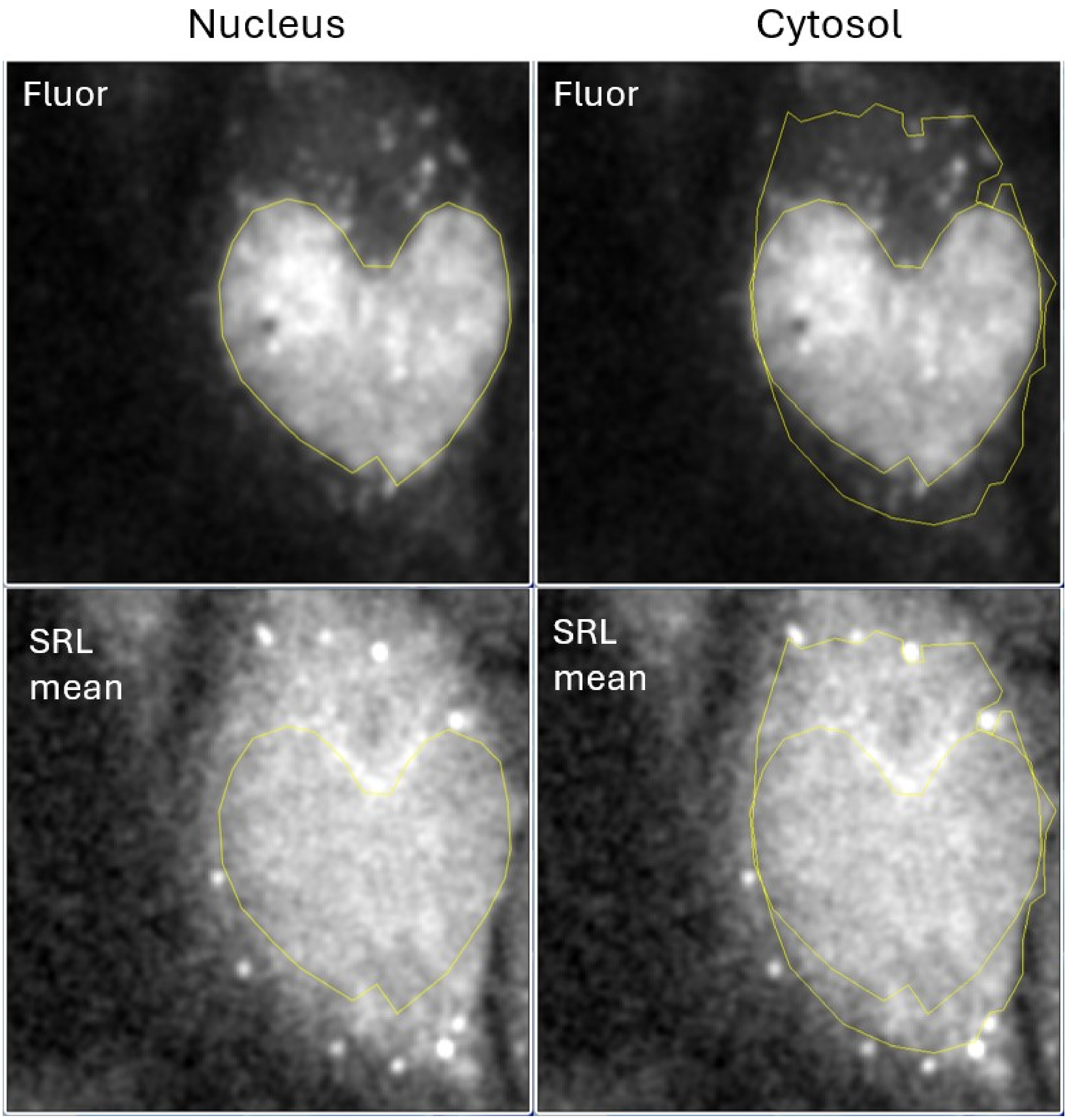

**Figure S12. Regions of interest for data shown in Fig.3.** Left: fluorescent nuclear region containing mRhubarb720. Right: weakly fluorescent cytosol region used as reference. Data shown taken for IFD 1660 cm$^{-1}$.

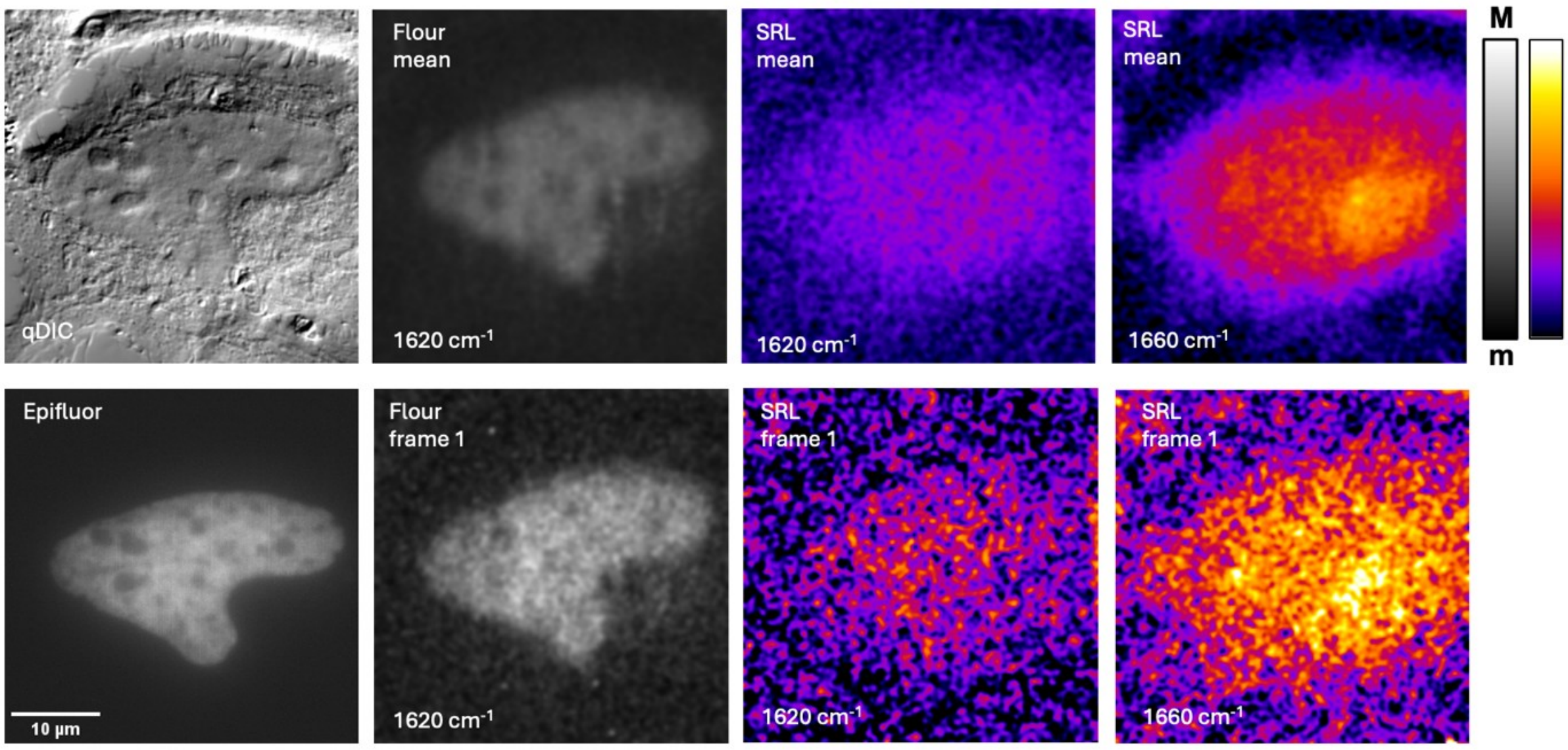

**Figure S13**: Imaging HeLa cell H2B-labelled mRhubarb720 expression. qDIC, SRL, widefield epi-fluorescence (Epifluor) and forward detected fluorescence (Fluor) are shown, as indicated. Grey and color scales are linear from m to M as follows: qDIC (m = 0.7 rad to M = 0.7 rad), Epifluor (m = 0 pe to M = 4 kpe, where pe are photoelectrons), Fluor (m = 0 Mpe/s to M = 1 Mpe/s, SRL (m = 0.2 mV to M = 1.3 mV). SRL and Fluor are taken over 40 frames, in alternating between 1620 and 1660 $cm^{-1}$, 0.01 ms pixel dwell time, 40×40 μm at 371×371 points, about 2s per frame. qDIC and Epifluor is taken before CRS imaging. The mean is taken over the 20 frames using the wavenumber given. Microscope objective 60× 1.27 NA, pump (Stokes) power at the sample 7 mW (12 mW). Centre IFD 1640 $cm^{-1}$.

## Section S7 One and Two-photon absorption and fluorophore symmetry

Epr-SRS requires close to resonant light excitation of a strong dipole-allowed transition (typically via the pump beam, while the Stokes beam is more red-shifted). The transition energy without vibrational excitation can be seen as the reflection symmetry point between emission and absorption spectra (see spectrum in Fig. S15 for Rhodamine 800). As discussed in the main text, red-shifting the light excitation compared to this point reduces the one-photon absorption by the thermal occupation factor. At room temperature, this provides 4 to 5 orders of magnitude reduction for 2000 $cm^{-1}$ shift. The fluorophores used here have a peak molar extinction around $\varepsilon$ = 100 $mM^{-1}$ $cm^{-1}$, corresponding to an absorption cross-section of $3.8 \times 10^{-16}$ $cm^2$.

For typical excitation powers of 10 mW at 100 MHz repetition rate at a wavelength of 820 nm focussed to the diffraction limit of a high numerical aperture objective (0.1 $\mu m^2$) e.g. our 1.27 NA objective, a peak photon flux of about $4\times10^{17}$ $cm^{-2}$ per pulse is created. Therefore, to have a one-photon absorption probability per pulse well below unity, an absorption cross-section of well below $2.4\times10^{-18}$ $cm^2$ is required. Considering the detuning of the pump beam at 820 nm from the transition without vibrational excitation around 715 nm for mRhubarb720, a 1790 $cm^{-1}$ shift is estimated, creating a thermal reduction factor of around 5000, which is consistent with the measured absorption (see Fig. S7). For the measured peak extinction coefficient of $\varepsilon$ = 72.5 $mM^{-1}$ $cm^{-1}$ (see table S1), the corresponding absorption cross-section is $5\times10^{-20}$ $cm^2$ at 820 nm, yielding a one-photon absorption probability in the focus per pump pulse of 0.02. This is well below unity, but still is providing a significant excitation, namely an excitation rate of 2 MHz. The corresponding fluorescence is clearly seen in the volume experiment in Fig. S6 comparing non-confocal with confocal detection.

Two-photon absorption is a mechanism which is always relevant as a background for CRS, as it is also a third-order non-linear signal, so that both CRS and TPA scale similarly with excitation pulse parameters and require high-excitation intensities. Assuming 1 ps pulse duration and the above-mentioned pulse parameters, the peak photon flux in the focus is $4\times10^{29}$ $cm^{-2}s^{-1}$. The two-photon cross-sections $\sigma_2$ of fluorophores depend strongly on their symmetry, being proportional to the change of the permanent dipole moment between ground and excited state.[19] Typical values of $\sigma_2$ are 10-100 GM ($1GM=10^{-50}$ $cm^4s$), from which we can calculate 0.016 to 0.16 absorbed photons during the 1 ps pulse considered above. Therefore, we find that one and two-photon absorption are on the same order of magnitude here, thus both contributing to photobleaching. Any larger detuning of the pump to reduce the one-photon absorption would reduce the epr-CRS while likely not reducing the excitation of the fluorophore by two-photon absorption. It is thus clear that TPA is in general a significant factor limiting epr-SRS, by providing background signal and creating photobleaching after the excitation. SRS background is created by TPA of the modulated Stokes beam, either by itself, with the fluorophore creating gain for the pump beam, or as Stokes-pump TPA directly attenuating the pump. The latter is clearly dominating as the background signal has the same sign as SRL. This background is visible in the pump-Stokes delay dependence that we have in spectral focussing (see e.g. Fig. S10) and is discussed in [20] for non-spectral focussing SRS.

TPA can occur at the frequencies $2\nu_p$, $\nu_p+\nu_S$ , and $2\nu_S$, which correspond to wavelengths of 410nm, 442nm, and 480 nm, respectively, for the wavelengths of 820 nm (pump) and 960 nm (Stokes) used in our experiment. These wavelengths are shown in Fig. S7, and we see that the pump TPA is spectrally matching with the second absorption band around 400 nm. TPA spectra of iRFP670 and iRFP720 have been reported in [21], although without quantification, as shown in Fig. S14. iRFP720 is spectrally similar to mRhubarb720, with excitation and emission wavelengths equal within 1 nm, and extinction coefficients equal within a few percent (see www.fpbase.org/protein/irfp720/ and www.fpbase.org/protein/mrhubarb720/). However, iRFP720 is a dimer, making it more bulky and thus less suited for genetic labelling. The spectra are consistent with the expectation that the highest TPA is occurring for the pump.

In this respect, it is interesting to discuss the SRL signal to background (S/B) ratios for a range of fluorophores reported in ref [22] (extended data table 1). The S/B ratio is defined as the vibrationally resonant signal relative to the non-resonant signal. The measurements by Wei et al[22] used a Stokes wavelength of 1064 nm and a pump wavelength in the range 860-920 nm, with 6 ps pulses at 80 MHz repetition rate without spectral focussing. In Fig. S15 we show dyes from this data with absorption peaks in the 680 to 710 nm region, selected to approximately retain the detuning from the resonance and thus supress detuning effects on the S/B ratio. Despite this, there are large variations of the reported S/B ratio. Notably, the dyes of high S/B (Rhodamine 800, Atto 700, MARS 2183) have delocalised conjugated bonds of high symmetry and structural rigidity formed by connected rings, while dyes of low S/B have a lower symmetry and conjugated bonds delocalised over a linear chain. This is consistent with the creation of the electronic background by two-photon absorption of the $S_0$-$S_1$ transition, which is stronger in less symmetric structures. The spectral shape of the two-photon absorption also plays a role. There are 3 sum frequencies available, $2\nu_p$, $\nu_p+\nu_S$ , and $2\nu_S$, which in the reported data correspond to wavelengths of 430-460 nm, 476-494 nm, and 532 nm, respectively. Two-photon absorption spectra of these fluorophores are not available to our knowledge.

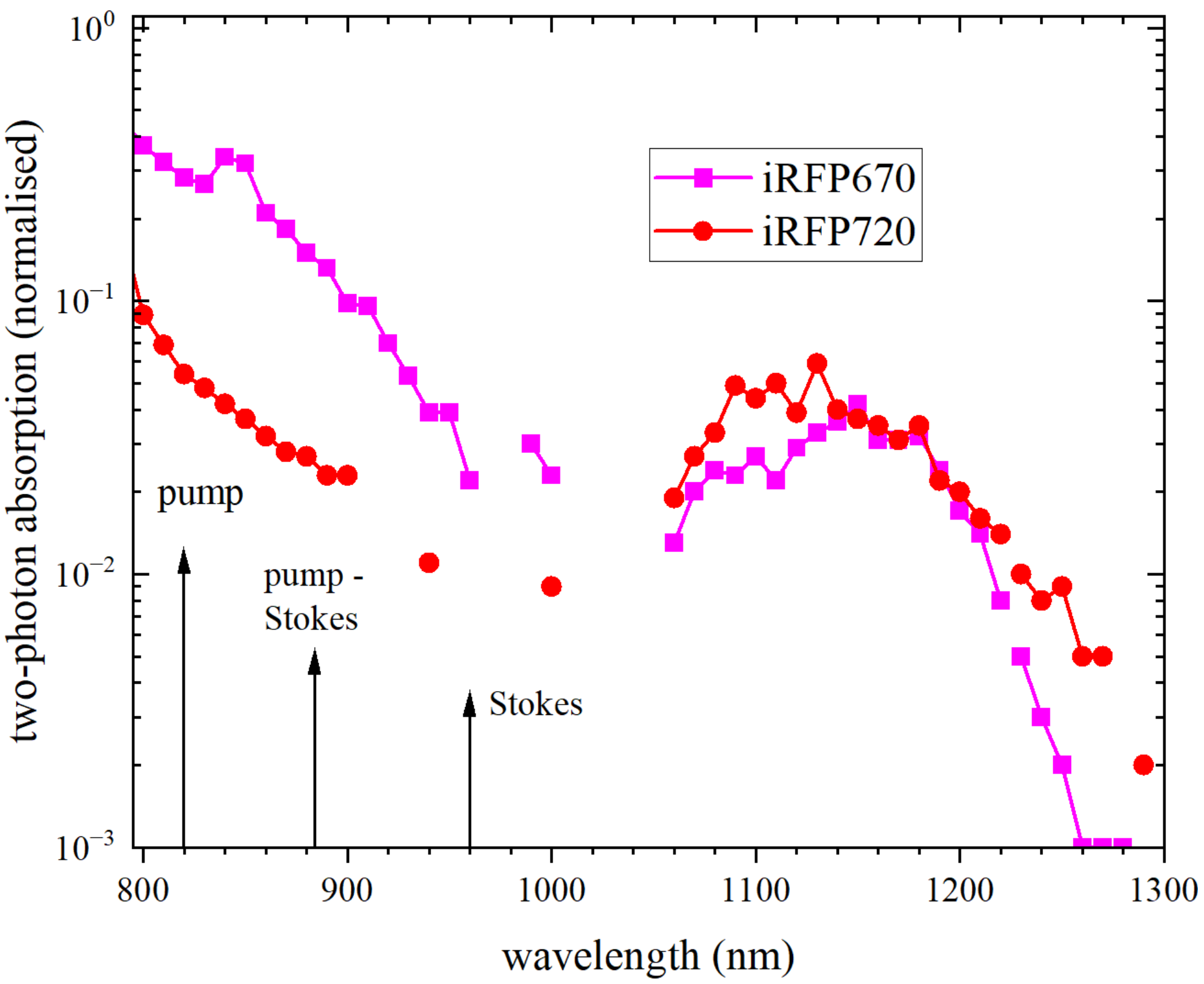

**Figure S14.** Two-photon absorption of iRFP670 and iRFP720 as reported previously.[21] Data is given normalised in arbitrary units. The pump, Stokes and pump-Stokes wavelength used in the present work are shown as arrows.

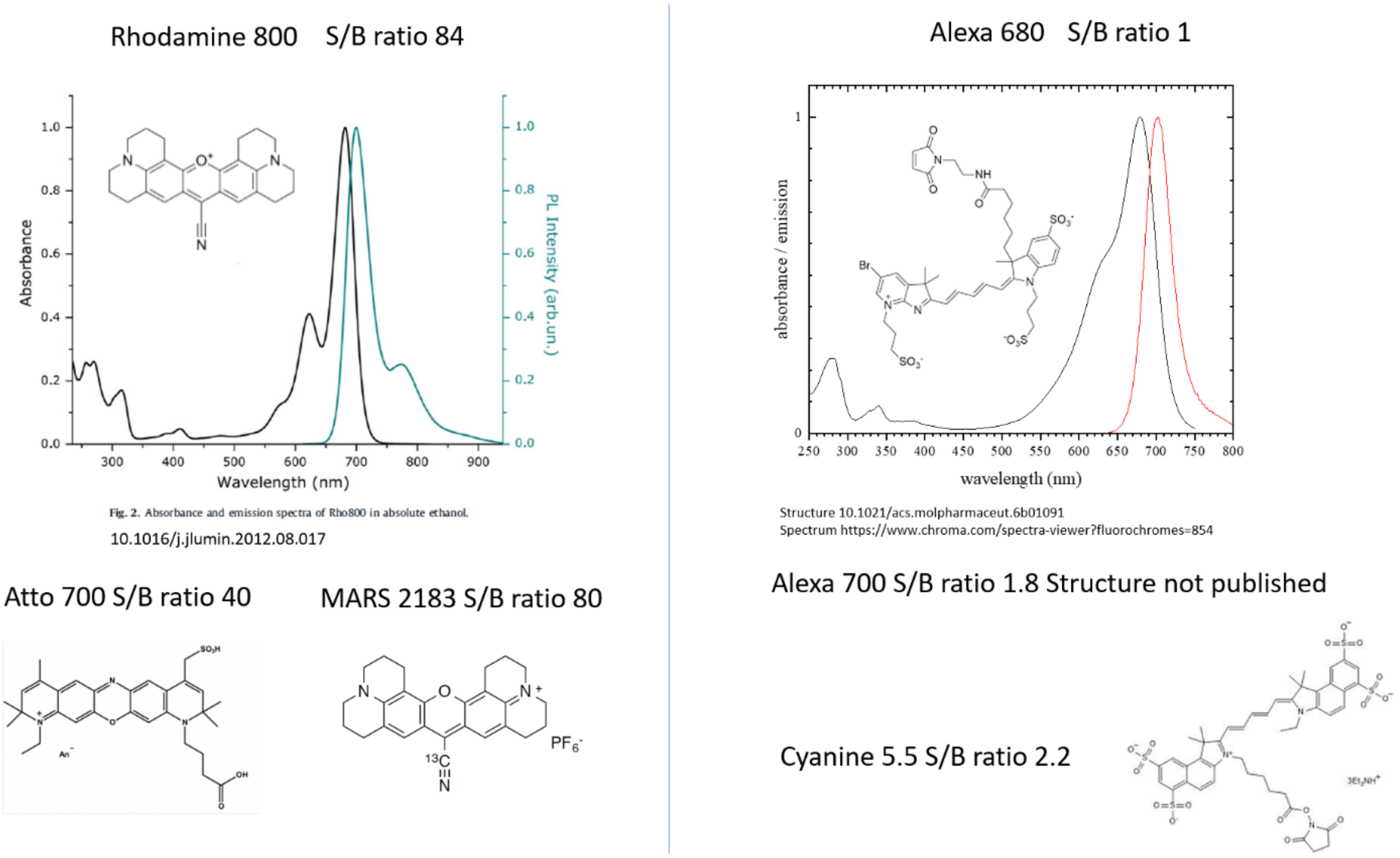

**Figure S15.** Signal to background (S/B) ratio for selected dyes reported by Wei et al[22]. The S/B ratio is defined as resonant vibrational contrast to non-resonant background in SRL using a Stokes wavelength of 1064nm and a Pump wavelength in the range 860-920nm. The absorption and emission spectra are given for Rhodamine 800 and Alexa 680. All dyes shown have the absorption peak in the 680-710 nm range.

## Section S8 Photobleaching during SRS imaging

For the imaging conditions in Fig.3 and S13, with a pixel dwell time of 0.01ms and a pixel size of 108 nm, about a third of the focus size, the exposure time for a given image point (consisting in 2D of 3x3=9 pixels) is about 0.1 ms per frame. The observed bleaching in Fig. 3 appears on a timescale of 10 frames, hence about 1 ms exposure. Assuming an excitation probability of the order of 10% per pulse as estimated in section S7, this corresponds to about $10^4$ excitations. We note that typical fluorophores photobleach after some $10^4$ to $10^6$ excitation cycles for the commonly used continuous wave excitation in their main absorption band. In our pulsed train excitation, there are 800 pulses during the pixel dwell time, which enables excited state absorption by subsequent pulses once an excitation has been created and still present after 12ns when the next pulse arrives. Fluorescence lifetimes of mRhubarb720 are around 1ns, therefore only intersystem crossing into the long-lived triplet state would allow the excitation to be still present when the next pulse arrives. This might be the main damage mechanism. Notably, since this absorption is spectrally broad, the absorption of pump and Stokes could be significant, exaggerating the photobleaching. Using a lower repetition rate laser and/or faster scanning to apply only a few pulses to any given position would enable reduction of such photobleaching.

An interesting aspect of the pre-resonant excitation is the stimulated emission of an excited fluorophore by the pump and Stokes. The stimulated emission can be expected to be much stronger than the one-photon absorption due to the red-shifted gain spectrum (given by the fluorescence spectrum), as also used in stimulated emission depletion (STED) microscopy.[23,24]

Thus, after an excitation, the stimulated emission is expected to remove the excitation efficiently, if there is sufficient time for vibrational relaxation. This could reduce photobleaching. However, as is well known from STED, at the same time, excited state absorption leads to an increase in photobleaching.

**Section S9 SRS detection sensitivity**

The results shown in Fig.2a at 1640cm-1 centre IFD, after subtracting the TPA contribution (blue solid line), show a peak SRS signal of 5.8mV. These data were acquired at 1mM concentration of mRhubarb720, using a pump power of 6mW and a Stokes power of 15mW (unmodulated) at the sample, respectively. The DC SRS voltage during these measurements was 79.3 mV. Considering the detector DC transimpedance of 36 Ohm, this corresponds to $I$=2.2mA DC current. The shot-noise in the experiment is calculated as root mean square (rms) current fluctuation given by $\sigma=\sqrt{2qI\Delta f}$ where $q$ is the electron charge and $\Delta f$ is the frequency bandwidth. With $I$=2.2mA, this corresponds to 26.5pA/$\sqrt{\mathrm{Hz}}$ and in turn (26.5pA/2.2mA)/$\sqrt{\mathrm{Hz}}$ = $1.2\times10^{-8}$ /$\sqrt{\mathrm{Hz}}$ relative modulation. At 1ms dwell time $\Delta f$=500Hz. We detect the SRS current in one quadrature only, which reduces the noise by $1/\sqrt{2}$, resulting in a relative modulation of $1.9\times10^{-7}$. The signal of 5.8mV detected for 1mM concentration, corresponds to a current of 58nA, using the detector transimpedance of $10^5$ Ohm at the SRS resonant modulation (2.5MHz). Therefore, the detected signal corresponds to a relative modulation of $\Delta$I/I =58nA/2.2mA=$2.6 \times 10^{-5}$. Scaling the shot-noise limit relative modulation of $1.9 \times 10^{-7}$ to the value $2.6 \times 10^{-5}$ at 1mM, equates to a detection limit concentration of 1mM $\times$ ($1.9 \times 10^{-7}$)/($2.6 \times 10^{-5}$) =7.2µM. Previously Wei and Min [25] used 24mW pump and 120mW stokes power at the sample, respectively. The detected SRS signal scales linearly with pump and Stokes power, while the noise scales with the square root of the pump power. Therefore, scaling the 7.2µM concentration limit to these powers results in a detection limit of 7.2µM $\sqrt{6/24}$ x (15/120) = 452nM.

**Supporting References**

(1) Lambert, T. J. FPbase: A Community-Editable Fluorescent Protein Database. *Nat Methods* 2019, *16* (4), 277–278. https://doi.org/10.1038/s41592-019-0352-8.

(2) Rogers, O. C.; Johnson, D. M.; Firnberg, E. MRhubarb: Engineering of Monomeric, Red-Shifted, and Brighter Variants of IRFP Using Structure-Guided Multi-Site Mutagenesis. *Sci Rep* 2019, *9* (1). https://doi.org/10.1038/s41598-019-52123-7.

(3) Matlashov, M. E.; Shcherbakova, D. M.; Alvelid, J.; Baloban, M.; Pennacchietti, F.; Shemetov, A. A.; Testa, I.; Verkhusha, V. V. A Set of Monomeric Near-Infrared Fluorescent Proteins for Multicolor Imaging across Scales. *Nat Commun* 2020, *11* (1). https://doi.org/10.1038/s41467-019-13897-6.

(4) Oliinyk, O. S.; Baloban, M.; Clark, C. L.; Carey, E.; Pletnev, S.; Nimmerjahn, A.; Verkhusha, V. V. Single-Domain near-Infrared Protein Provides a Scaffold for Antigen-Dependent Fluorescent Nanobodies. *Nat Methods* 2022, *19* (6), 740–750. https://doi.org/10.1038/s41592-022-01467-6.

(5) Warren, A. J.; Trincao, J.; Crawshaw, A. D.; Beale, E. V.; Duller, G.; Stallwood, A.; Lunnon, M.; Littlewood, R.; Prescott, A.; Foster, A.; Smith, N.; Rehm, G.; Gayadeen, S.; Bloomer, C.; Alianelli, L.; Laundy, D.; Sutter, J.; Cahill, L.; Evans, G. VMXm – A Sub-Micron Focus Macromolecular Crystallography Beamline at Diamond Light Source. *J Synchrotron Radiat* 2024, *31* (6), 1593–1608. https://doi.org/10.1107/S1600577524009160.

(6) Winter, G.; Beilsten-Edmands, J.; Devenish, N.; Gerstel, M.; Gildea, R. J.; McDonagh, D.; Pascal, E.; Waterman, D. G.; Williams, B. H.; Evans, G. DIALS as a Toolkit. *Protein Science* 2022, *31* (1), 232–250. https://doi.org/10.1002/pro.4224.

(7) Gildea, R. J.; Beilsten-Edmands, J.; Axford, D.; Horrell, S.; Aller, P.; Sandy, J.; Sanchez-Weatherby, J.; Owen, C. D.; Lukacik, P.; Strain-Damerell, C.; Owen, R. L.; Walsh, M. A.; Winter, G. Xia2.Multiplex: A Multi-Crystal Data-Analysis Pipeline. *Acta Crystallogr D Struct Biol* 2022, *78* (6), 752–769. https://doi.org/10.1107/S2059798322004399.

(8) Jumper, J.; Evans, R.; Pritzel, A.; Green, T.; Figurnov, M.; Ronneberger, O.; Tunyasuvunakool, K.; Bates, R.; Žídek, A.; Potapenko, A.; Bridgland, A.; Meyer, C.; Kohl, S. A. A.; Ballard, A. J.; Cowie, A.; Romera-Paredes, B.; Nikolov, S.; Jain, R.; Adler, J.; Back, T.; Petersen, S.; Reiman, D.; Clancy, E.; Zielinski, M.; Steinegger, M.; Pacholska, M.; Berghammer, T.; Bodenstein, S.; Silver, D.; Vinyals, O.; Senior, A. W.; Kavukcuoglu, K.; Kohli, P.; Hassabis, D. Highly Accurate Protein Structure Prediction with AlphaFold. *Nature 2021 596:7873* 2021, *596* (7873), 583–589. https://doi.org/10.1038/s41586-021-03819-2.

(9) Collaborative Computational Project, N. 4. The CCP4 Suite: Programs for Protein Crystallography. *Acta Crystallogr D Biol Crystallogr* 1994, *50* (Pt 5), 760–763. https://doi.org/10.1107/S0907444994003112.

(10) Emsley, P.; Cowtan, K. Coot: Model-Building Tools for Molecular Graphics. *Acta Crystallogr D Biol Crystallogr* 2004, *60* (Pt 12 Pt 1), 2126–2132. https://doi.org/10.1107/S0907444904019158.

(11) Baloban, M.; Shcherbakova, D. M.; Pletnev, S.; Pletnev, V. Z.; Lagarias, J. C.; Verkhusha, V. V. Designing Brighter Near-Infrared Fluorescent Proteins: Insights from Structural and Biochemical Studies. *Chem Sci* 2017, *8* (6), 4546–4557. https://doi.org/10.1039/C7SC00855D.

(12) Shaner, N. C.; Campbell, R. E.; Steinbach, P. A.; Giepmans, B. N. G.; Palmer, A. E.; Tsien, R. Y. Improved Monomeric Red, Orange and Yellow Fluorescent Proteins Derived from Discosoma Sp. Red Fluorescent Protein. *Nat Biotechnol* 2004, *22* (12), 1567–1572. https://doi.org/10.1038/nbt1037.

(13) Langbein, W.; Regan, D.; Pope, I.; Borri, P. Invited Article: Heterodyne Dual-Polarization Epi-Detected CARS Microscopy for Chemical and Topographic Imaging of Interfaces. *APL Photonics* 2018, *3* (9). https://doi.org/10.1063/1.5027256.

(14) Langbein, W.; Rocha-Mendoza, I.; Borri, P. Coherent Anti-Stokes Raman Micro-spectroscopy Using Spectral Focusing: Theory and Experiment. *Journal of Raman Spectroscopy* 2009, *40* (7), 800–808. https://doi.org/10.1002/jrs.2264.

(15) Albrecht, A. C.; Hutley, M. C. On the Dependence of Vibrational Raman Intensity on the Wavelength of Incident Light. *J Chem Phys* 1971, *55* (9), 4438–4443. https://doi.org/10.1063/1.1676771.

(16) Xiong, H.; Min, W. Combining the Best of Two Worlds: Stimulated Raman Excited Fluorescence. *J Chem Phys* 2020, *153* (21). https://doi.org/10.1063/5.0030204.

(17) Hamilton, S.; Regan, D.; Payne, L.; Langbein, W.; Borri, P. Sizing Individual Dielectric Nanoparticles with Quantitative Differential Interference Contrast Microscopy. *Analyst* 2022, *147* (8), 1567–1580. https://doi.org/10.1039/D1AN02009A.

(18) Regan, D.; Williams, J.; Borri, P.; Langbein, W. Lipid Bilayer Thickness Measured by Quantitative DIC Reveals Phase Transitions and Effects of Substrate Hydrophilicity. *Langmuir* 2019, *35* (43), 13805–13814. https://doi.org/10.1021/acs.langmuir.9b02538.

(19) Drobizhev, M.; Makarov, N. S.; Tillo, S. E.; Hughes, T. E.; Rebane, A. Two-Photon Absorption Properties of Fluorescent Proteins. *Nat Methods* 2011, *8* (5), 393–399. https://doi.org/10.1038/nmeth.1596.

(20) Pruccoli, A.; Kocademir, M.; Winterhalder, M. J.; Zumbusch, A. Electronically Preresonant Stimulated Raman Scattering Microscopy of Weakly Fluorescing Chromophores. *J Phys Chem B* 2023, *127* (27), 6029–6037. https://doi.org/10.1021/acs.jpcb.3c01407.

(21) Leben, R.; Lindquist, R. L.; Hauser, A. E.; Niesner, R.; Rakhymzhan, A. Two-Photon Excitation Spectra of Various Fluorescent Proteins within a Broad Excitation Range. *Int J Mol Sci* 2022, *23* (21), 13407. https://doi.org/10.3390/ijms232113407.

(22) Wei, L.; Chen, Z.; Shi, L.; Long, R.; Anzalone, A. V.; Zhang, L.; Hu, F.; Yuste, R.; Cornish, V. W.; Min, W. Super-Multiplex Vibrational Imaging. *Nature* 2017, *544* (7651), 465–470. https://doi.org/10.1038/nature22051.

(23) Vicidomini, G.; Bianchini, P.; Diaspro, A. STED Super-Resolved Microscopy. *Nat Methods* 2018, *15* (3), 173–182. https://doi.org/10.1038/nmeth.4593.

(24) Hell, S. W.; Wichmann, J. Breaking the Diffraction Resolution Limit by Stimulated Emission: Stimulated-Emission-Depletion Fluorescence Microscopy. *Opt Lett* 1994, *19* (11), 780. https://doi.org/10.1364/OL.19.000780.
(25) Wei, L.; Min, W. Electronic Preresonance Stimulated Raman Scattering Microscopy. *J Phys Chem Lett* 2018, *9* (15), 4294–4301. https://doi.org/10.1021/acs.jpclett.8b00204.